\documentclass[onecolumn,showpacs,preprint,amsmath,amssymb,superscriptaddress]{revtex4-1}

\usepackage{dcolumn}
\usepackage{bm}
\usepackage{hyperref}
\newcommand{\ignore}[1]{}
\usepackage{graphicx}
\usepackage{amsmath,amssymb,amsfonts,amsthm,color,latexsym}

\usepackage[utf8]{inputenc}
\usepackage[T1]{fontenc}
\usepackage{mathptmx}

\usepackage{soul}

\usepackage{tabstackengine}

\usepackage[utf8]{inputenc}
\usepackage[T1]{fontenc}
\usepackage{mathptmx}

\def\mb{\mathbf}

\def\bm{\boldsymbol}
\newcommand{\Rmnum}[1]{\uppercase\expandafter{\romannumeral #1\relax}}
\newcommand{\rmnum}[1]{\lowercase\expandafter{\romannumeral #1\relax}}
\def\mb{\mathbf}
\newcommand*\diff{\mathop{}\!\mathrm{d}}

\mathchardef\mhyphen="2D
\usepackage{amsmath}

\usepackage[normalem]{ulem}

\usepackage[export]{adjustbox}

\begin{document}
\preprint{}

\title{Machine learning assisted coarse-grained molecular dynamics modeling of meso-scale interfacial fluids}
\author{Pei Ge}
\affiliation{Department of Computational Mathematics, Science \& Engineering, Michigan State University, East Lansing, MI 48824, USA}%
\author{Linfeng Zhang}
\thanks{linfeng.zhang.zlf@gmail.com}
\affiliation{AI for Science Institute, Beijing 100080, China}%
\affiliation{DP Technology, Beijing 100080, China}%
\author{Huan Lei}
\thanks{leihuan@msu.edu}
\affiliation{Department of Computational Mathematics, Science \& Engineering, Michigan State University, East Lansing, MI 48824, USA}%
\affiliation{Department of Statistics \& Probability, Michigan State University, East Lansing, MI 48824, USA}%


\begin{abstract}
A hallmark of meso-scale interfacial fluids is the multi-faceted, scale-dependent interfacial energy, which often manifests different characteristics across the molecular and continuum scale. The multi-scale nature imposes a challenge to construct reliable coarse-grained (CG) models, where the CG potential function needs to faithfully encode the many-body interactions arising from the unresolved atomistic interactions and account for the heterogeneous density distributions across the interface.  We construct the CG models of both single$\mhyphen$ and two$\mhyphen$component of polymeric fluid systems based on the recently developed deep coarse-grained potential (DeePCG) \cite{Zhang_DeePCG_JCP_2018} scheme, where each polymer molecule is modeled as a CG particle. By only using the training samples of the instantaneous force under the thermal equilibrium state, the constructed CG models can accurately reproduce both the probability density function of the void formation in bulk and the spectrum of the capillary wave across the fluid interface. More importantly, the CG models accurately predict the volume$\mhyphen$to$\mhyphen$area scaling transition for the apolar solvation energy, illustrating the effectiveness to probe the meso-scale collective behaviors encoded with molecular-level fidelity.


\end{abstract}


\maketitle
\section{Introduction}

Molecular dynamics (MD) simulations provide a promising avenue to establish the atomistic-level understanding of 
many complex systems relevant to biological and materials science. Despite the overwhelming success during the past decades, a remaining bottleneck roots in the limitation of the achievable spatio-temporal scales; the gap between the micro-scale atomistic motions and many meso-scale emerging phenomena remains large. One important problem is the nano-scale interfacial fluids, which play a crucial role in the hydration and the assembly of the biomolecules and functional nano-materials \cite{Chandler:2005ds, Weeks:2009dg}. However, it is well-known that such fluid systems generally exhibit complex and multifaceted nature on different scales. On the small scale (i.e., the fluid molecule correlation length), the solvation energy is determined by the molecular reorganization and scales with the volume of the void space. On the large scale, the solvation energy is determined by the free energy for maintaining a fluid-void interface and scales with the surface area. The scale-dependent behavior indicates an cross-over regime of the entropy-enthalpy transition. While theoretical understandings \cite{Lum_Chandler_JPCB_1999, Rein_Chandler_PRE_2001, Hummer_Pratt_PNAS_1996, Hummer_Pratt_JPCB_1998} of this ubiquitous phenomenon have been developed, computational modeling often relies on full micro-scale MD simulations to retain the multifaceted properties, which, however, remain too expensive to achieve the resolved scale for applications such as nano-scale assembly.

To accelerate the full MD simulations, many coarse-grained (CG) models have been developed. 
By modeling the dynamics in terms of a set of CG variables with reduced dimensionality, the coarse-grained molecular dynamics (CGMD) simulations, in principle, enable us to probe the collective behaviors on a broader scale. 
However, in practice, the construction of truly reliable CG models can be highly non-trivial, especially for the meso-scale interfacial fluids. There are two major challenges. The first challenge arises from the many-body nature of CG interactions. Specifically, the equilibrium density distribution of the CG model needs to match the marginal density distribution of the CG variables of the full model. Due to the unresolved atomistic degrees of freedom, the CG potential generally encodes the many-body interactions even if the full MD force field is governed by two-body interactions \cite{noid_multiscale_2008}. Existing approaches often rely on various physical intuitions as well as empirical approximations \cite{Izvekov_Voth_JPC_2005, noid_perspective_2013,Lei_Cas_2010, hijon2010mori, Rudd_APS_1998, Pagonabarraga_JCP_2001,Nielsen_JPCM_2004,shinoda2008coarse,Molinero_JPCB_2009,Larini_JCP_2010,das2012multiscale,dinpajooh2017density,sanyal2016coarse,moore2016coarse} that reproduce certain target thermodynamic quantities and/or structural distributions. For example, the pairwise additive decomposition based on direct ensemble averaging \cite{Lei_Cas_2010,hijon2010mori} can recover the thermodynamic pressure but often fail to recover the pair distribution function.   
Conversely, the Monte Carlo and Boltzmann inverse approaches \cite{Lyubartsev_PRE_1995, Soper_Chem_Phys_1996, Reith_JCP_2003} can reproduce the pairwise distribution function, which, however, lead to the biased predictions of the equation of state. Several studies account for the many-body effects by introducing the configuration-independent volume potential \cite{Das_And_JCP_2010,Dunn_Noid_JCP_2015,Dunn_Noid_JCP_2016} and the local density \cite{Allen_Rut_JCP_2008,Allen_Rut_JCP_2009,Izvekov_Chung_JCP_2010,moore2016coarse,sanyal2016coarse,Shahidi_Chazirakis_JCP_2020} into the pairwise interactions. On the other hand, the accuracy of the high-order structural correlations as well as the direct applications to interfacial systems remains under-explored. 

Besides the many-body effect, the fluid molecules also exhibit heterogeneous density at the interfacial vicinity. What further complicates the problem is the fact that the interfacial fluid density distribution is \emph{scale-dependent}. On the small scale, the molecular reorganization generally leads to a wet interface with larger density than the bulk value. On the large scale, the fluid-void phase separation generally leads to a dry interface with lower density. The crossover implies complex molecular correlations near the interface. To capture this multi-faceted property, the constructed CG potential needs to properly embody the local particle distribution other than the homogeneous bulk distribution. Conventional structural-based CG potential functions generally show limitations to incorporate such information. Similar to the many-body dissipative particle dynamics \cite{Pagonabarraga_JCP_2001}, recent studies employed the local density \cite{wagner2017extending, delyser2017extending,Sanyal_Shell_JPCB_2018,Jin_Voth_JCTC_2018,DeLyser_Noid_JCP_2019,DeLyser_Noid_JCP_2020,Berressem_Scherer_JPCM_2021} as well as the density gradient \cite{DeLyser_Noid_JCP_2022} as the auxiliary field variables to construct the CG potential functions. While the CG models show significant improvement to reproduce the interfacial density profile, the scale-dependent interfacial energy and fluctuations have not been systematically investigated. In Ref. \cite{Lei_Schenter_JCP_2015}, interfacial energy is integrated into the continuum fluctuation hydrodynamic equation \cite{Landau_Lifshitz_Fluid_1987} from the top-down perspective. Fluid particles essentially represent the Lagrangian discretization points based on the smoothed dissipative particle hydrodynamics \cite{Serrano_Espanol_PRE_2001} instead of the CG molecules; the meso-scale fluid structural properties can not be retained. Currently, the construction of reliable bottom-up CGMD models that faithfully encode the multifaceted molecular interactions remains largely open.

In this work, we aim to address the above challenges by constructing CG models of meso-scale interfacial fluids based on the deep molecular dynamics (DeePMD) scheme \cite{PRL_DeePMD_2018,Zhang_BookChap_NIPS_2018_v31_p4436}.
DeePMD is initially developed for learning the many-body interactions from the \emph{ab initio} molecular dynamics, and has been applied to construct the deep coarse-grained (DeePCG) model \cite{Zhang_DeePCG_JCP_2018} of liquid water in bulk. Unlike the conventional forms of the inter-molecular potential function, the DeePMD represents each particle as an agent and the relative positions of its neighboring particles as the local environment. Rather than approximating the total potential of the full system by an unified parametric function, the DeePMD directly maps the local environment of each agent to the potential energy of that particle through a neural network that strictly preserves the spatial symmetries and the particle permutation invariance. Accordingly, the construction does not rely on the empirical  decomposition (e.g., pairwise, three-body) of the high-dimensional particle configuration space. This unique feature is particularly suited for modeling the many-body potential of CGMD models, where the ensemble-averaged interaction between two CG particles further depends on the other neighboring CG particles and can not be represented by a pairwise additive function. Moreover, the heterogeneous particle density distribution across the fluid interface can be naturally incorporated into the CG potential function as the local environment of each particle. Accordingly, the constructed CG models can accurately model the multifaceted, scale-dependent interfacial fluctuations and apolar solvation without additional human intervention. 

We demonstrate the effectiveness of the CG models by considering both the single$\mhyphen$ and two-component fluids in presence of thermal interfacial fluctuations. As discussed in Ref. \cite{Chandler:2005ds}, the scale-dependent hydrophobic effects can be general for solvent molecules with attractive interactions; polymeric liquids are therefore used as the benchmark problem. We compare the numerical results from the full MD simulations and the CG description that represents each molecule as a single particle located at the center of mass.
By merely using training samples under equilibrium thermal fluctuations, the constructed CG models accurately predict the high-order correlations, the local compressibility, and the interfacial capillary wave. In contrast, the empirical CG potential constructed based on the pairwise approximation shows apparent deviations. More importantly, the CG models accurately predict the probability of void formation in bulk as well as the volume-to-area scaling transition for the solvation energy, and therefore, pave the way for modeling the nanoscale assembly in aqueous environment. 

Before wrapping up this section, we note that the present work focuses on the collective, quasi-equilibrium properties determined by the conservative potential function of a set of extensive CG variables; see Refs. \cite{John_ST_JPCB_2017, ML_CG_water_Nature_2019} for relevant work.  For the conformational free energy of non-extensive CG variables, several machine-learning based approaches \cite{Stecher_Thomas_JCTC_2014, Mones_Letif_JCTC_2016, Lemke_Tobias_JCTC_2017, Galvelis_Raimondas_JCTC_2017,Schneider_Elia_APS_2017,Zhang_Linfeng_JCP_2018,Zavadlav_Julija_JCTC_2018,Noe_Clementi_ACS_2019} have been developed; see also a recent review \cite{Noe_Tka_ARPC_2020} and the references therein. Furthermore, to accurately predict the dynamic properties, memory and coherent noise terms \cite{Mori1965, Zwanzig73} arising from the unresolved variables need to be properly introduced into the CG model \cite{Lei_Cas_2010, hijon2010mori,Lei_Li_PNAS_2016, Lei_Li_JCP_2021}, which are left to future investigations.

\section{Methods and models}

\subsection{Full model of the polymeric fluids}
\label{sec:full_model}
We consider the micro-scale models of the star polymer melt similar to Ref. \cite{hijon2010mori}. The full system consists of $M$ molecules with a total number of $N$ atoms. Each polymer molecule consists of a ``center'' atom connected by
$N_a$ arms with $N_b$ atoms per arm. The positions of the atoms are denoted by $\mb q = \left[\mb q_1, \mb q_2, \cdots, \mb q_N\right]$, where $\mb q_i$ represents the position of the $i\mhyphen$th atom. The potential function is governed by the pairwise and bond interactions, i.e.,
\begin{equation}
V(\mb q) = \sum_{i\neq j} V_{p}(q_{ij}) + \sum_k V_{b}(l_k),
\label{eq:MD_potential}
\end{equation}
where $V_{p}$ is the pairwise interaction between both the intra- and inter-molecular atoms except the bonded pairs. $q_{ij} = \Vert \mb q_i - \mb q_j\Vert$ 
is the distance between the $i\mhyphen$th and $j\mhyphen$th atoms.  
$V_{b}$ is the bond interaction between the neighboring  particles of each polymer arm and $l_k$ is the length of the $k\mhyphen$th bond. 
The bond potential $V_b$ is chosen to be the harmonic potential, i.e.,
\begin{equation}
V_{b}(l) = \frac{1}{2}k_s (l - l_0)^2, 
\label{eq:harmonic}
\end{equation}
where $k_s$ and $l_0$ represent the elastic coefficient and the equilibrium length $l_0$, respectively. The atom mass is chosen to be unity.

We investigate three fluid systems with micro-scale potential governed by Eq. \eqref{eq:MD_potential}. In Sec. \ref{sec:bulk_fluids}, we consider the polymeric fluids in bulk and examine if the CG models can retain the many-body interactions and the local compressibility. 
In particular, we choose $N_a = 12$, $N_b = 6$, $\sigma = 2.415$, $\epsilon = 1.0$, $k_s = 1.714$, $l_0 = 2.77$ similar to Ref. \cite{hijon2010mori}.  $V_{p}$ takes the form of the Lennard–Jones potential with cut-off $r_c$, i.e., 
\begin{equation}
V_{\rm p}(r) = \begin{cases} V_{\rm LJ}(r) - V_{\rm LJ}(r_c), ~r < r_c  \\
0, ~r \ge r_c 
\end{cases} \quad
\quad V_{\rm LJ}(r) = 4\epsilon \left[\left(\frac{\sigma}{r}\right)^{12} -  \left(\frac{\sigma}{r}\right)^6\right],
\label{eq:LJpotential}
\end{equation}
where $\epsilon = 1.0$ is the dispersion energy and $\sigma = 2.415$ is the hardcore distance. Also we choose $r_c = 2^{1/6}\sigma$ so that $V_p$ recovers the Weeks-Chandler-Andersen potential. 
The full system consists of $N = 2120$ polymer molecules in a cubic domain $180\times180\times180$ with periodic boundary condition imposed along each direction. The Nos\'{e}-Hoover thermostat is employed to conduct the canonical ensemble simulation with $k_BT =  3.96$. 

In Sec. \ref{sec:one_component_fluids}, we consider the polymeric fluid in presence of fluid-void interface. Micro-scale model parameters are similar to Sec. \ref{sec:bulk_fluids} except that $r_c = 2.5\sigma$ and $k_BT = 1.7$. Simulations are conducted in a domain $180\times180\times200$ with periodic boundary condition imposed along the $x\mhyphen$ and $y\mhyphen$direction. At the equilibrium, the fluid shows a clear fluid-void interface near $z = 20$ and $z = 180$, respectively. 

In Sec. \ref{sec:two_component_fluids}, we consider a two-component polymeric fluid. Micro-scale model of the polymer molecule is similar to the single-component fluid system with $N_a = 15$, $N_b = 12$, $k_s = 20.0$, $l_0 = 1.5$, $k_BT =  0.5$. The full system consists of 3488 molecules in a domain $200\times200\times120$ with periodic boundary condition imposed along each direction. The pairwise interaction $V_p$ is chosen to be quadratic, i.e.,
\begin{equation}
V_{\rm p}(r) = \begin{cases} \frac{a}{2r_c}\left(r - r_c\right)^2, r < r_c \\
0, r \ge r_c 
\end{cases}.
\end{equation}
Specifically, we consider two sets of the pairwise interaction: (\Rmnum{1}) $\sigma^{11} = 6.0$, $\sigma^{12} = 3.0$, $\sigma^{22} = 6.0$, $r_c = 1.5$ , where $\sigma^{12}$ represents the pairwise interaction between the component-1 and component-2 atoms. (\Rmnum{2}) $\sigma^{11} = 3.0$, $\sigma^{12} = 60.0$, $\sigma^{11} = 3.0$, $r_c^{11} = 1.5$, $r_c^{12} = 2.5$, $r_c^{11} = 1.5$.  The fluid shows a full mixture and interfacial separated state for the two cases respectively.

\subsection{Coarse-grained models}

For all of the three systems, we construct the CG models by representing each molecule as an individual particle. The positions of the CG particles are denoted by $\mb Q =  \left[\mb Q_1, \mb Q_2, \cdots, \mb Q_M\right]$, where $\mb Q_i = \mathcal{Q}_i(\mb q)$ represents the center of mass (COM) of the $i\mhyphen$th molecule. The conservative
potential $U(\mb Q)$ is determined by the marginal density function of $\mb Q$ with respect to the equilibrium density function of the full model, i.e., 
\begin{equation}
\begin{split}
   \rho (\mb Q) &=\int e^{-V(\mb q)/k_BT} \prod_{i=1}^M \delta(\mathcal{Q}_i(\mb q)-\mb Q_i) \diff \mb q / \int e^{-V(\mb q)/k_BT} \diff \mb q, \\
   U(\mb Q)&=-k_BT\ln \rho (\mb Q).
\end{split}
\label{eq:rho_U_Q}
\end{equation}

In DeePCG, a neural network $\tilde{U}(\mb Q; \bm\Theta)$ is used to represent the CG potential $U(\mb Q)$, where $\bm\Theta$ represents the neural network parameters. To keep the extensive property, the total energy is decomposing into local contributions of the individual CG particles:
\begin{equation}
\tilde{U}(\mb Q;\theta)=\sum_{i=1}^M \tilde{U}_{nn}(\mb D(\tilde{\mb Q}^i); \theta), 
\label{eq:DeeP_CG}
\end{equation}
where $\tilde{\mb Q}^i \in \mathbb{R}^{N_i \times 4}$ is the generalized coordinates of the $i\mhyphen$th particle. It represents the local environment of the $i\mhyphen$th particle relative to its $N_i$ neighboring particles within cutoff $R_c$. In particular, the $j\mhyphen$th row is defined as $ \tilde{\mb Q}^i_j=(s(r_{j}), s(r_j)x_j/r_j^2, s(r_j)y_j/r_j^2, s(r_j)z_j/r_j^2)$, where $\mb r_j = (x_j, y_j, z_j)$ denotes the relative position between the $i\mhyphen$th particle and its $j\mhyphen$th local neighbor. $s(r)$ is a smooth differentiable function that decays to $0$ at $r=R_c$. $\mb D \in R^{M_1 \times M_2}$ is the symmetry preserving features of each particle. 
Each entry of $D$ can be written as:
\begin{equation}
D_{j,l}(\tilde{\mb Q}^i)=\left(\sum_{k=1}^{N_i} g_{1,j}(s(r_k)) \tilde{\mb Q}^i_k \right)\left(\sum_{k=1}^{N_i} g_{2,l}(s(r_k)) \tilde{\mb Q}^i_k\right)^T,
\end{equation}
where $\left\{g_{1,j}(r)\right\}_{j=1}^{M_1}$ and $\left\{g_{2,l}(r)\right\}_{j=1}^{M_2}$ are neural networks mapping from the scalar $r$ to multiple features. $D_{j,l}$ preserves the translational and rotational invariance; the summation over index $k$ ensures the permutational symmetry.

In principle, $\tilde{U}(\mb Q;\bm\Theta)$ can be trained by minimizing the difference of the predicted force terms between  the full micro-scale and the CG models, i.e., $\left \langle \Vert \nabla \tilde{U}(\mb Q;\bm\Theta) -  \nabla U(\mb Q) \Vert^2\right\rangle_{\mb Q}$, where $\left\langle \cdot\right\rangle_{\mb Q}$ represents the conditional expectation with respect to the constraints of $\mb Q$, i.e., $ \prod_{i=1}^M \delta(\mathcal{Q}_i(\mb q)-\mb Q_i)$. However, the evaluation of the force term $-\nabla U(\mb Q)$ relies on the constraint sampling with respect to $\delta(\mathcal{Q}(\mb q)-\mb Q)$, which can be computational expensive. On the other hand, we note that the instantaneous force $\mathcal{F}(\mb Q)$ follows $\mathcal{F}(\mb Q) = -  \nabla U(\mb Q) + \mathcal{R}(\mb Q)$, where $\mathcal{R}(\mb Q)$ is the zero-mean fluctuation force. Therefore, we have 
$\left \langle \Vert \nabla \tilde{U}(\mb Q;\bm\Theta) - \nabla U(\mb Q)   \Vert^2 \right \rangle_{\mb Q} = \left \langle \Vert \nabla \tilde{U}(\mb Q;\bm\Theta) + \mathcal{F}(\mb Q)\Vert^2 \right \rangle_{\mb Q} +  \left \langle \Vert \mathcal{R}(\mb Q) \Vert^2 \right\rangle_{\mb Q}$, where the last term does not involve in the training. Accordingly, we can transform the training by minimizing the empirical loss 
\begin{equation}
L = \sum_{i=1}^S \sum_{j=1}^M \left\Vert \nabla \tilde{U}(\mb Q^{(i)};\bm\Theta) + \mathcal{F}_j(\mb Q^{(i)})\right\Vert^2,   
\end{equation}
where the superscript represents the index of $S$ configurations. For the three micro-scale models specified in Sec. \ref{sec:full_model}, we collect training samples from $50\mhyphen$, $200\mhyphen$, $250\mhyphen$long (in reduced unit) trajectories from the full MD simulations. $5000$ snapshots are used to train the CG potential function for each case. The networks are trained by the Adam stochastic gradient descent method \cite{Kingma_Ba_Adam_2015}. In particular, we emphasize that \emph{all the training samples are collected from thermal equilibrium states}. As shown in Sec. \ref{sec:numerical_results}, the constructed CG potentials naturally encode the many-body and heterogeneous interfacial interactions, which enable us to accurately predict rare events such as the probability of the void formation and scale-dependent apolar solvation energy.

\section{Numerical results}
\label{sec:numerical_results}
\subsection{Bulk fluids}
\label{sec:bulk_fluids}
Let us start with the CG model of fluids in bulk. Due to the constraint terms in Eq. \eqref{eq:rho_U_Q}, the marginal probability density function $\rho(\mb Q)$ generally can not be represented in form of the simple two-point correlation $\rho^{(2)}(\mb Q_i, \mb Q_j)$. Accordingly, the CG potential function $U(\mb Q)$ generally exhibits the many-body nature and can not be exactly constructed in form of the pairwise interaction. This limitation was verified in earlier studies on the CG modeling of polymeric fluids \cite{Lei_Cas_2010, hijon2010mori}, where the CG interactions are constructed based on the pairwise decomposition, i.e.,  
\begin{equation}
\begin{split}
U(\mb Q) &\approx \sum_{i\neq j} U^{(2)}(Q_{ij}) \\
\frac{\diff U^{(2)}(r)}{\diff r} &= -\left \langle \mathcal{F}_{ij}(\mb Q_{ij})  \cdot \mb e_{ij}\right\rangle_{Q_{ij} = r},    
\end{split}
\label{eq:CG_pairwise}
\end{equation}
where $\mb e_{ij} = \mb Q_{ij}/Q_{ij}$ represents the unit vector between the $i\mhyphen$th and $j\mhyphen$th particle. 

\begin{figure}[htbp]
\centering
\includegraphics[scale=0.32]{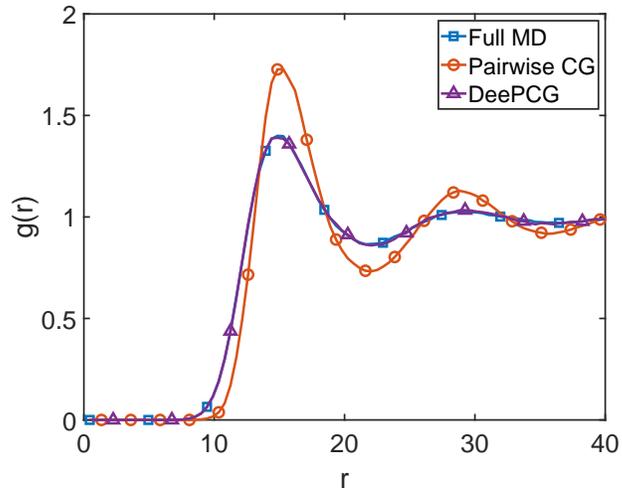}
\caption{Radial distribution function $g(r)$ of the molecule COM obtained from the full MD simulation, the CG model using the pairwise force approximation by Eq. \eqref{eq:CG_pairwise}, and the DeePCG model.}
\label{fig:rdf_polymer_melt_CG}
\end{figure}

To examine the model accuracy, we simulate the CG models with $U(\mb Q)$ constructed in form of both Eq. \eqref{eq:DeeP_CG} and Eq. \eqref{eq:CG_pairwise}. Fig. \ref{fig:rdf_polymer_melt_CG} shows
the obtained radial distribution functions (RDFs). Predictions from the full MD and the reduced model based on the DeePCG potential \eqref{eq:DeeP_CG} show good agreement. In contrast, the pairwise CG potential \eqref{eq:CG_pairwise} yields pronounced over-estimations of the peak value near $r = 16$ due to the over-simplification of the many-body CG potential using the two-body interaction; see also Refs. \cite{Lei_Cas_2010,hijon2010mori}. 

\begin{figure}[htbp]
\centering
\includegraphics[scale=0.24]{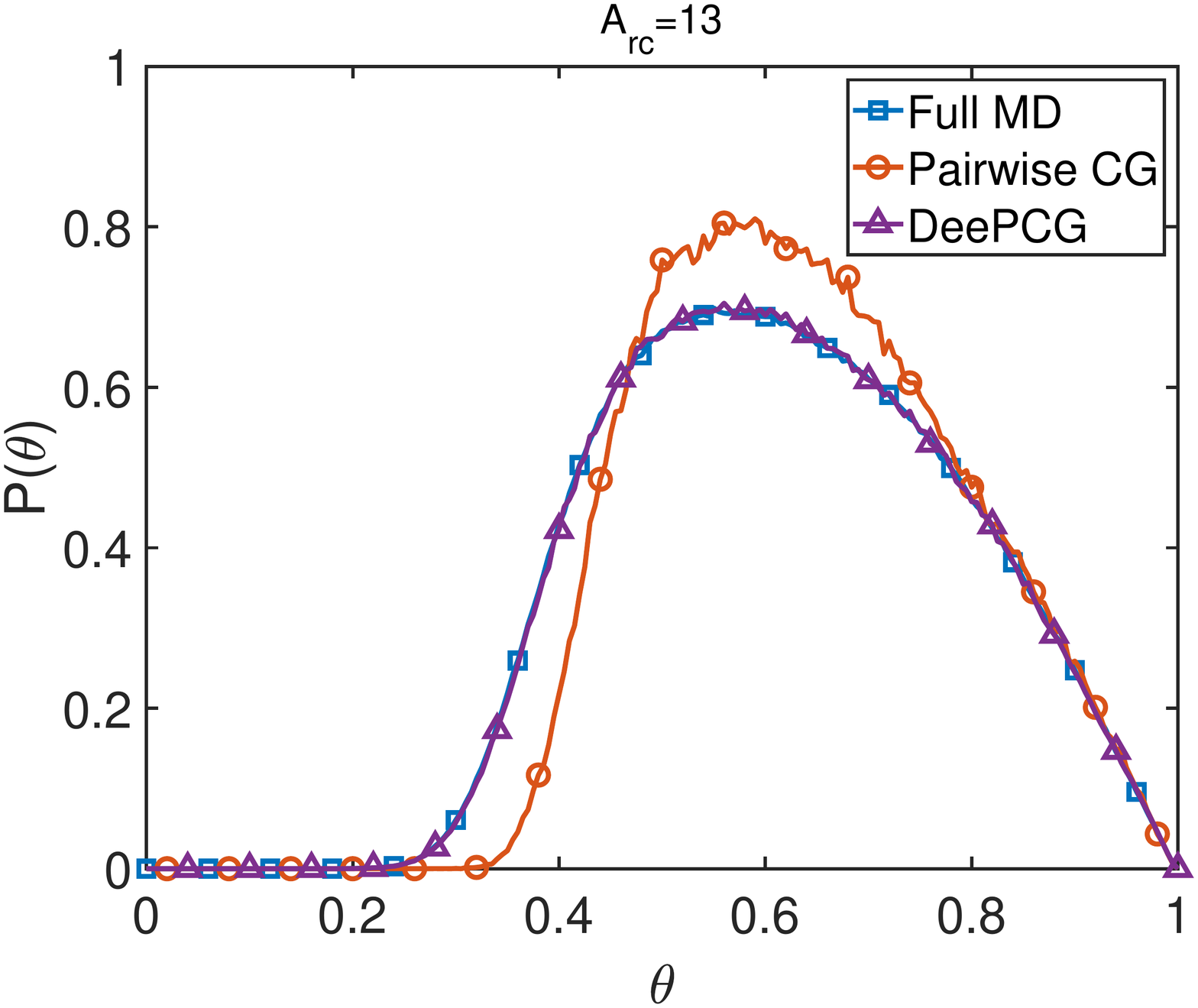}
\includegraphics[scale=0.24]{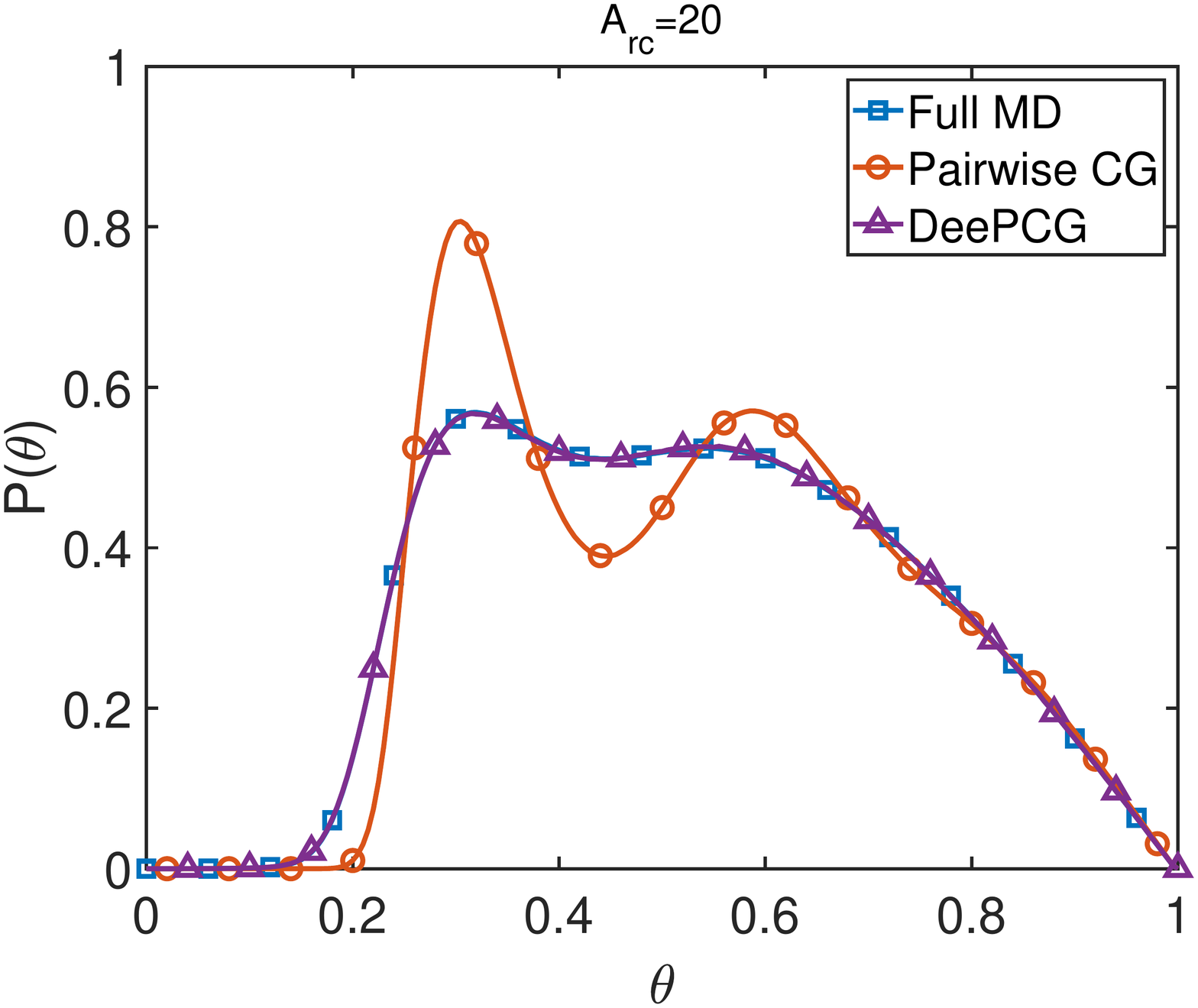}
\includegraphics[scale=0.24]{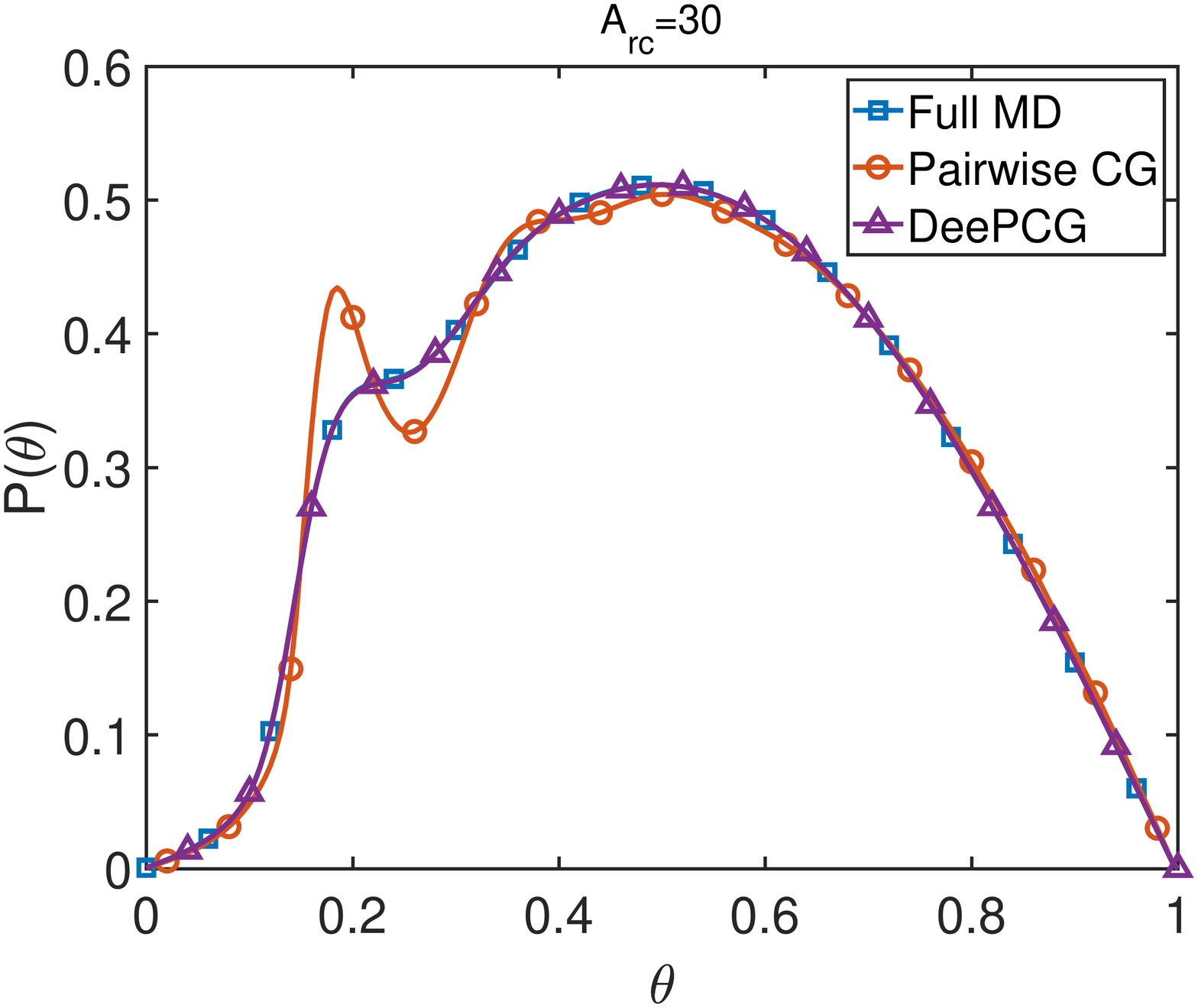}
\includegraphics[scale=0.24]{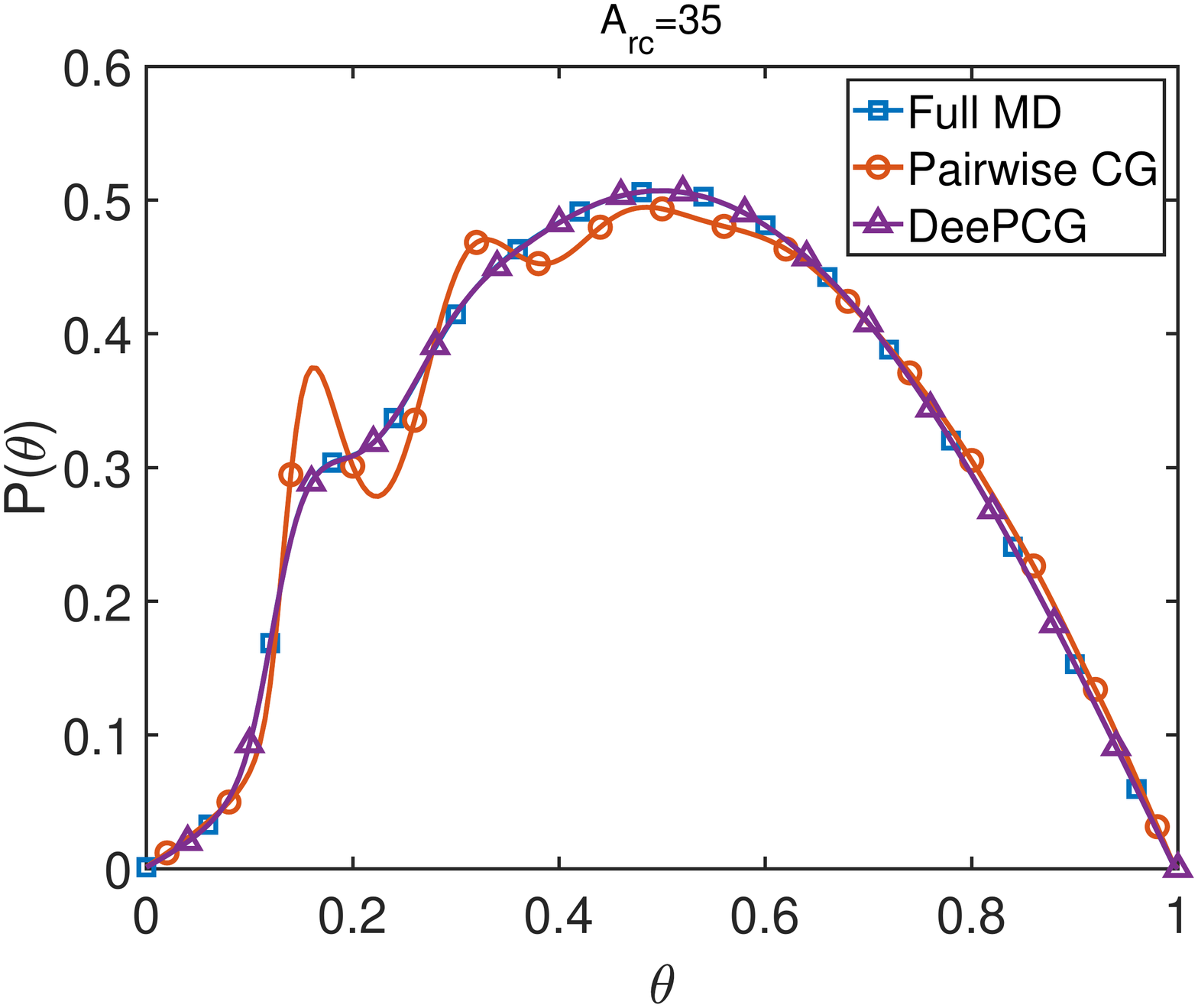}
\caption{Angular distribution function $p(\theta)$ of the molecule COM obtained from the full MD simulation, the pairwise CG model and the DeePCG model with different cut-off regimes $A_{rc}$.}
\label{fig:ADF_polymer_melt_CG}
\end{figure}

The many-body nature of $U(\mb Q)$ is also manifested in the angular distribution functions (ADFs) $p(\theta)$, where $\theta$ is the angle determined by relative positions of three molecules. 
\begin{equation}
    P(\theta;A_{rc})=\frac{1}{W}\left\langle \sum_i \sum_{j\neq i} \sum_{k>j}\delta(\theta-\theta_{jik}) \right\rangle
\end{equation}
where $\theta_{jik}$ is the angle between $\mb Q_{ji}$ and $\mb Q_{ki}$, and $W$ is a normalization factor. The summation is over all the
triplet $i$, $j$, $k$, such that $\Vert \mb Q_i-\mb Q_j\Vert\leq A_{rc}$ and $\Vert \mb Q_i- \mb Q_k\Vert\leq A_{rc}$.  Fig. \ref{fig:ADF_polymer_melt_CG} shows the ADFs within four different cut-off regimes. Similar to the RDF, predictions of the DeePCG model agree well with the full MD model while the pairwise approximation yields apparent deviations.  

Besides the equilibrium correlations, we further examine the fluid local compressibility. While this property plays an important role in the nano-scale hydrophobicity, canonical solvation theories generally refer to the fluids at the proximity of the vapor-liquid coexistent phase. 
Here we examine this property of bulk fluids for the validation of the constructed many-body CG potential $\tilde{U}(\mb Q)$; the discussion of the apolar solvation energy is postponed to Sec. \ref{sec:one_component_fluids}. Specifically, we examine the rare event of the void formation in bulk. Following Ref. \cite{Patel_Chandler_JPCB_2010}, we define the smoothed molecule number within a probing spherical volume centered at $\mb Q_c$ by 
\begin{equation}
\hat{n} \left(\left\{\mb Q_i\right\}_{i=1}^{M}\right) = \sum_{i=1}^M \frac{1}{2}\left( 1 + 2\tanh\left(\frac{R - Q_i}{h}\right) \right),
\label{eq:CG_number}
\end{equation}
where $R$ is the radius of the probing sphere, $Q_i = \Vert \mb Q_i - \mb Q_c \Vert$ is the distance between the COM of molecule $i$ (or equivalently, the CG particle) and the spherical center, and $h = 1.0$ represents the smooth length.

\begin{figure}[htbp]
\begin{center}
\begin{tabular}{c c c c c c}
\textbf{\scriptsize (a)} & \includegraphics[scale=0.175, valign=t]{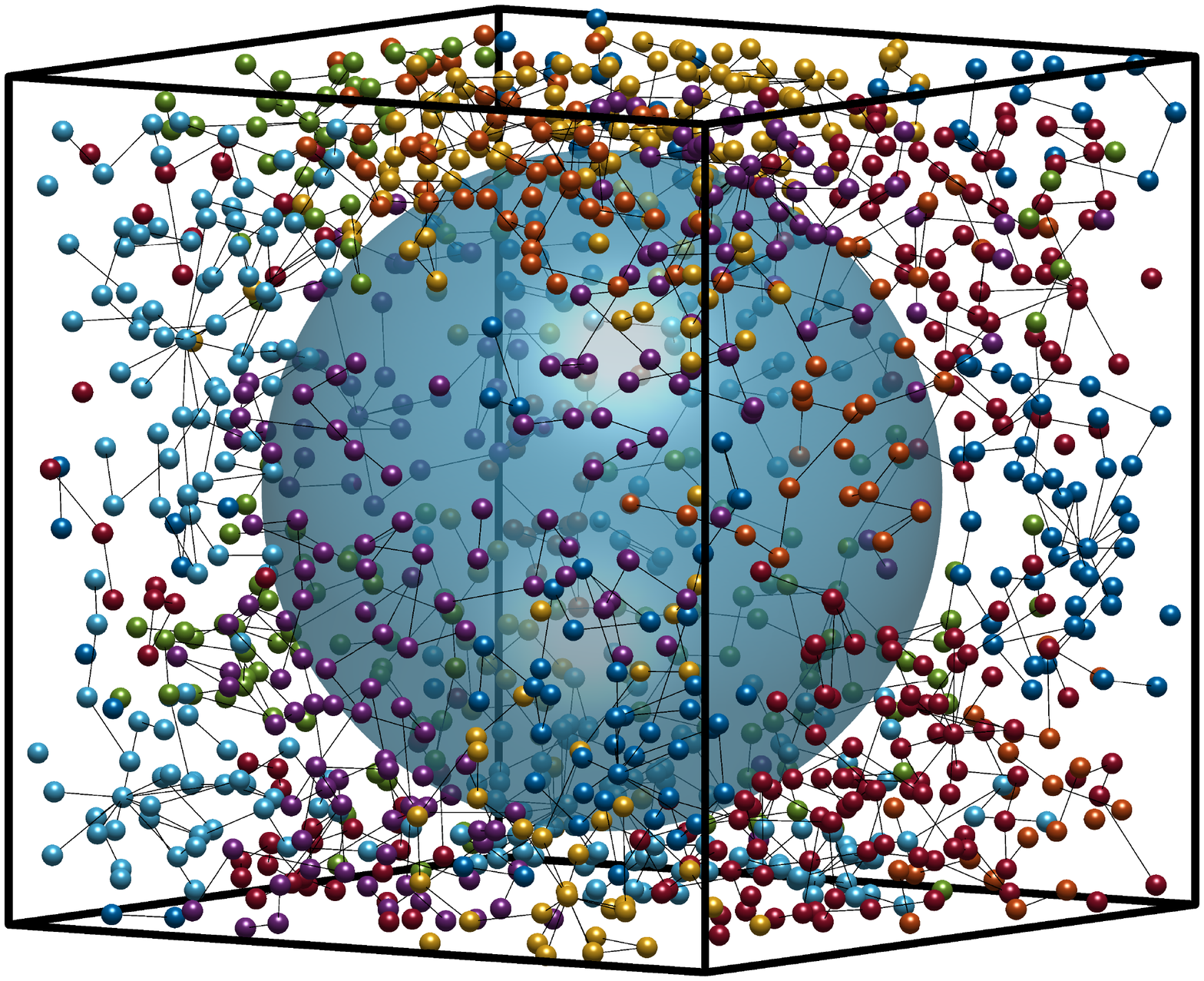}  & \textbf{\scriptsize (b)} & \includegraphics[scale=0.175, valign=t]{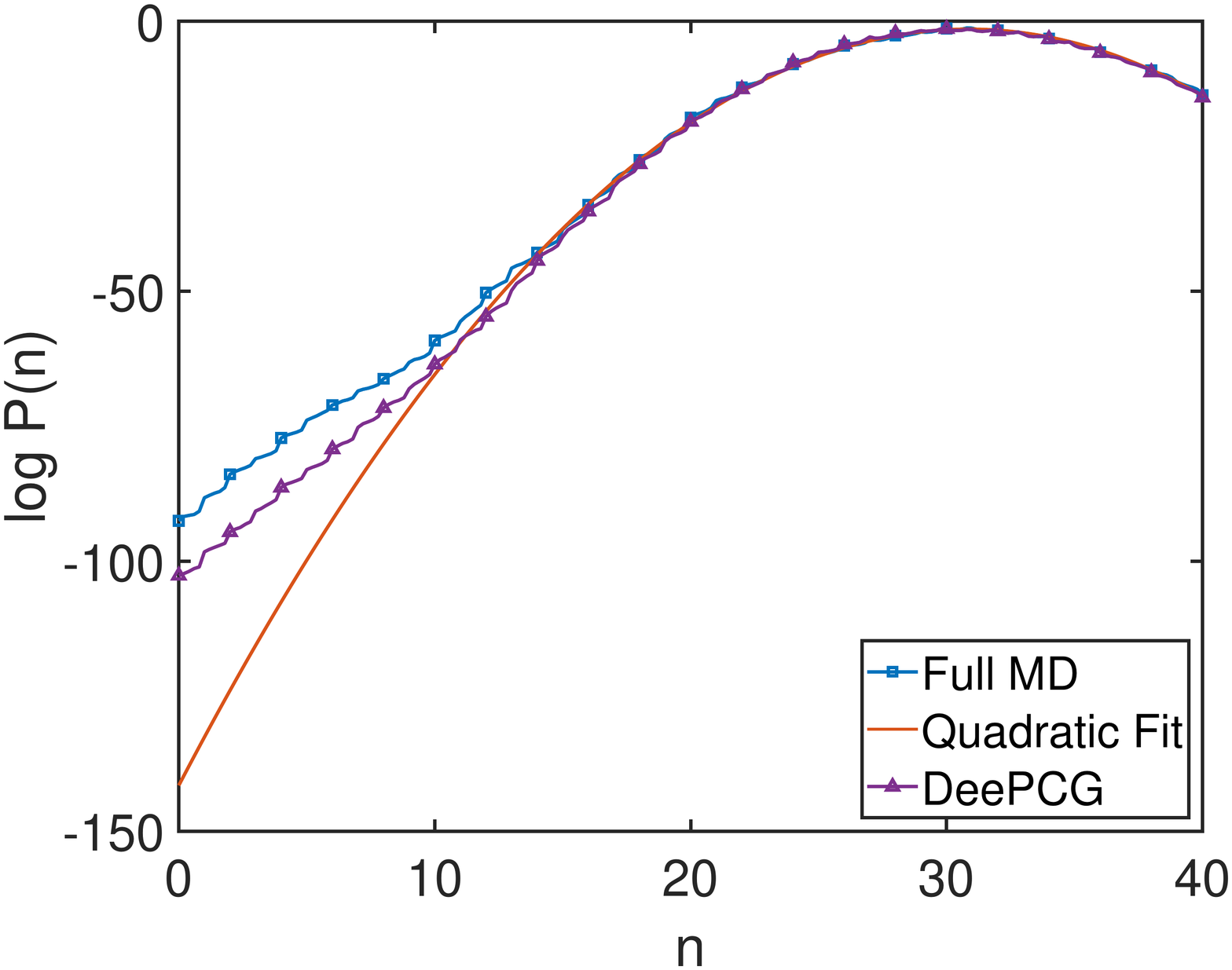} & \textbf{\scriptsize (c)} & \includegraphics[scale=0.175, valign=t]{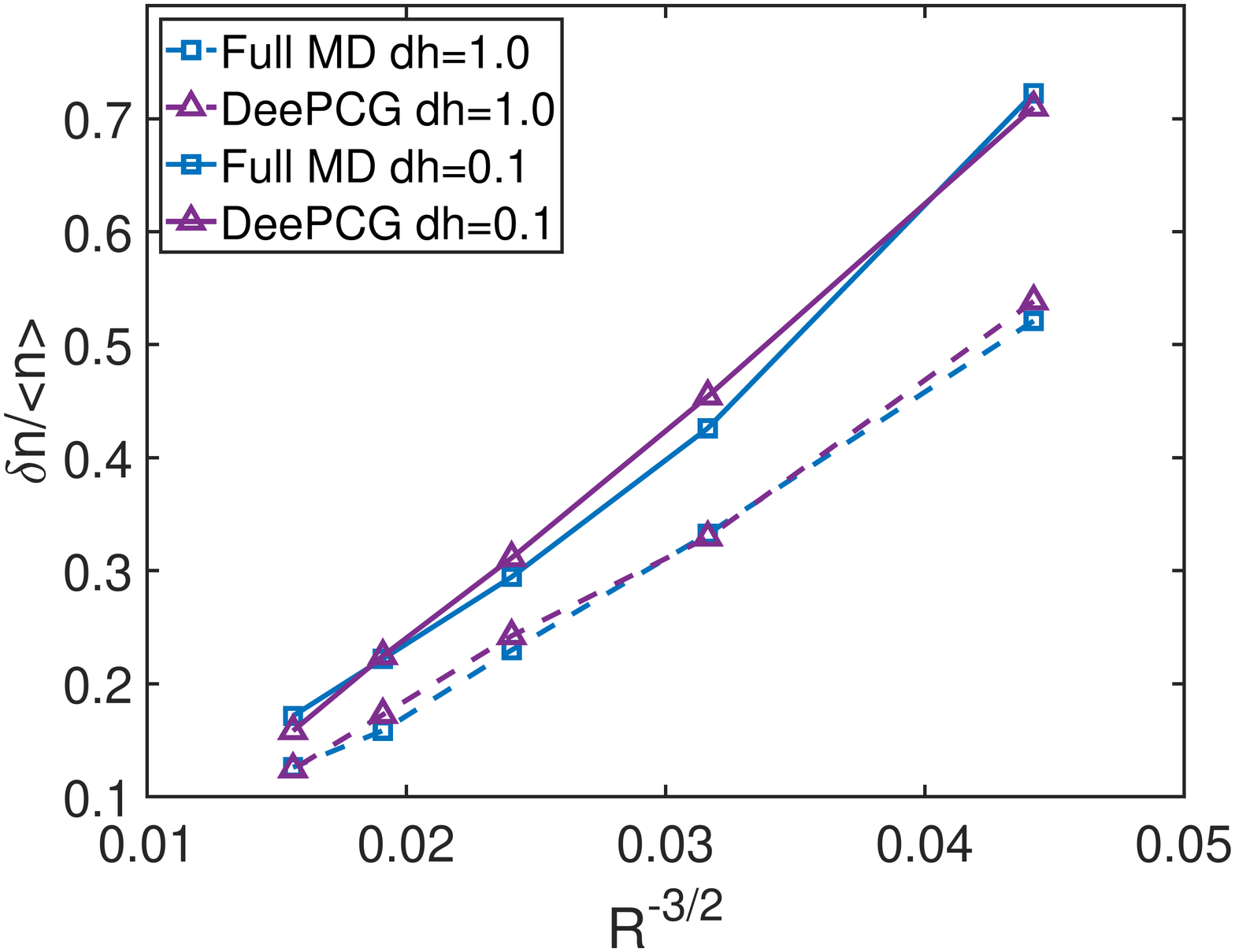}
\end{tabular}
\caption{The density fluctuation and the molecule number distribution within a spherical probing volume. (a) A sketch of the instantaneous molecule position
under bias potential \eqref{eq:bias_potential}. The iso-surface in blue color represents the interface of the void space. (b) The probability density function of the molecule number within a spherical volume of radius
$R = 16.0$. The vertical dashed line represents the average molecule number under equilibrium. (c) The normalized density fluctuations within a spherical
volume of radius $R$ between $8.0$ and $16.0$. The particle number is defined by Eq. \eqref{eq:CG_number} with the resolution length $h$ set
to be $0.1$ (solid lines) and $1.0$ (dashed lines).}
\label{fig:void_polymer_melt_CG}
\end{center}
\end{figure}

By Eq. \eqref{eq:CG_number}, particle number $\hat{n}$  is differentiable with respect to the 
individual molecule position $\mb Q_i$. Similar to Ref. \cite{Patel_Chandler_JPCB_2010}, we can probe the probability of the void formation by establishing a replica of umbrella sampling by imposing the bias potential
\begin{equation}
U_{\rm bias} (\hat{n}; n_j) = \frac{k_n}{2}(\hat{n} - n_j)^2,
\label{eq:bias_potential}
\end{equation}
where $k_n$ is the magnitude of the bias potential and $n_j$ is the target value of the particle number inside the domain, as shown in Fig. \ref{fig:void_polymer_melt_CG}(a). We set $k_n = 21.9$ and establish $40$ independent simulations with $n_j$ evenly distributed between $0$ and $7.5$. For each replica, we collect $8\times 10^5$ samples 
of $\hat{n}$ from a $1600$-long trajectory. By using the weighted histogram analysis method \cite{Kumar1992}, we can stitch the joint probability 
density  $\rho(\hat{n}, n_j)$ to construct $\rho(\hat{n})$. Fig. \ref{fig:void_polymer_melt_CG}(b) shows the probability density $\rho(\hat{n})$ obtained from the full MD and the reduced model. 
The predictions of the DeePCG model agree well with the full MD model over the full regime of $\hat{n}$.

Finally, we examine the normalized density fluctuation $\delta n/\langle n\rangle$ within a spherical volume of various sizes, where $\left\langle n\right\rangle$
is the average particle number and $\delta n = \sqrt{\left\langle \left(\hat{n} - \langle n \rangle\right)^2\right\rangle}$ is the standard deviation. 
Specifically, we define the particle number by Eq. \eqref{eq:CG_number} with two different smooth length $h = 1.0$ and $h = 0.1$, respectively. The latter case essentially 
represents each molecule as a \emph{simple point} and counts the particle number as integers, and therefore, yields larger density fluctuations. As shown in Fig. \ref{fig:void_polymer_melt_CG}(c), the full MD and CG model show good agreement for both cases, indicating that the CG model can faithfully capture the high-order correlations and the local compressibility beyond the continuum thermodynamic limit.


\subsection{Single-component interfacial fluids}
\label{sec:one_component_fluids}
Besides the many-body interactions, another hallmark of interfacial fluids is the heterogeneous molecular distribution across the fluid interface, which leads to scale-dependent interfacial interactions and fluctuations. On the macro-scale level, the interfacial interactions can be generally described by continuum models such as the Young-Laplace equation \cite{rowlinson2002molecular}; the apolar solvation energy is proportional to the interfacial area and characterized by the surface tensor. However, on the length scale comparable to the correlation length of the fluid molecules, the interfacial energy often exhibits a cross-over regime representing the volume-dependent to area-dependent scaling transition. Therefore, the meso-scale interfacial energy provides a crucial metric to validate the accuracy of the CG model.

First, we examine the interfacial thermal fluctuations. With the micro-scale model specified in Sec. \ref{sec:full_model}, the fluid molecule interaction consists of both the short-range repulsion and long-range attraction. Under the thermal equilibrium states, the fluid system exhibits the fluid-void interfaces near $z = 20$ and $z= 180$. The periodic boundary condition is imposed along the $x\mhyphen$ and $y\mhyphen$direction.

To quantify the molecule distribution near the interface at $z = z_0$, we define the smoothed density field $\rho_s(\mb R)$ by
\begin{equation}
\rho_s(\mb R) = \sum_{i=1}^M W\left(\Vert \mb R - \mb Q_i\Vert, h\right)
\label{eq:density_field}
\end{equation}
on the $N_x\times N_y\times N_z$ lattice grids. Specifically, $\mb R^{(i,j,k)} := (x^i, y^j, z^k)$, where $(x^i, y^j) = (i,j)\times dl$, $dl = L/N_x$ and 
$z^k = z_0 - h + k \times dz$, $dz = 2h/N_z$. 
$\mb Q_i$ represents the COMs of the neighboring molecules for each grid point. $W(r, h)$ represents the quintic spline kernel function \cite{Morris_Patrick_JCP_1997} with finite support $h$. In this study, we set $h = 30.0$, $dl = 1.8$ and $dz =0.2$.  

\begin{figure}[htbp]
\begin{center}
\begin{tabular}{ c c c c}
\textbf{\scriptsize (a)} & \includegraphics[scale=0.35, trim=120 100 120 120,clip, valign=t]{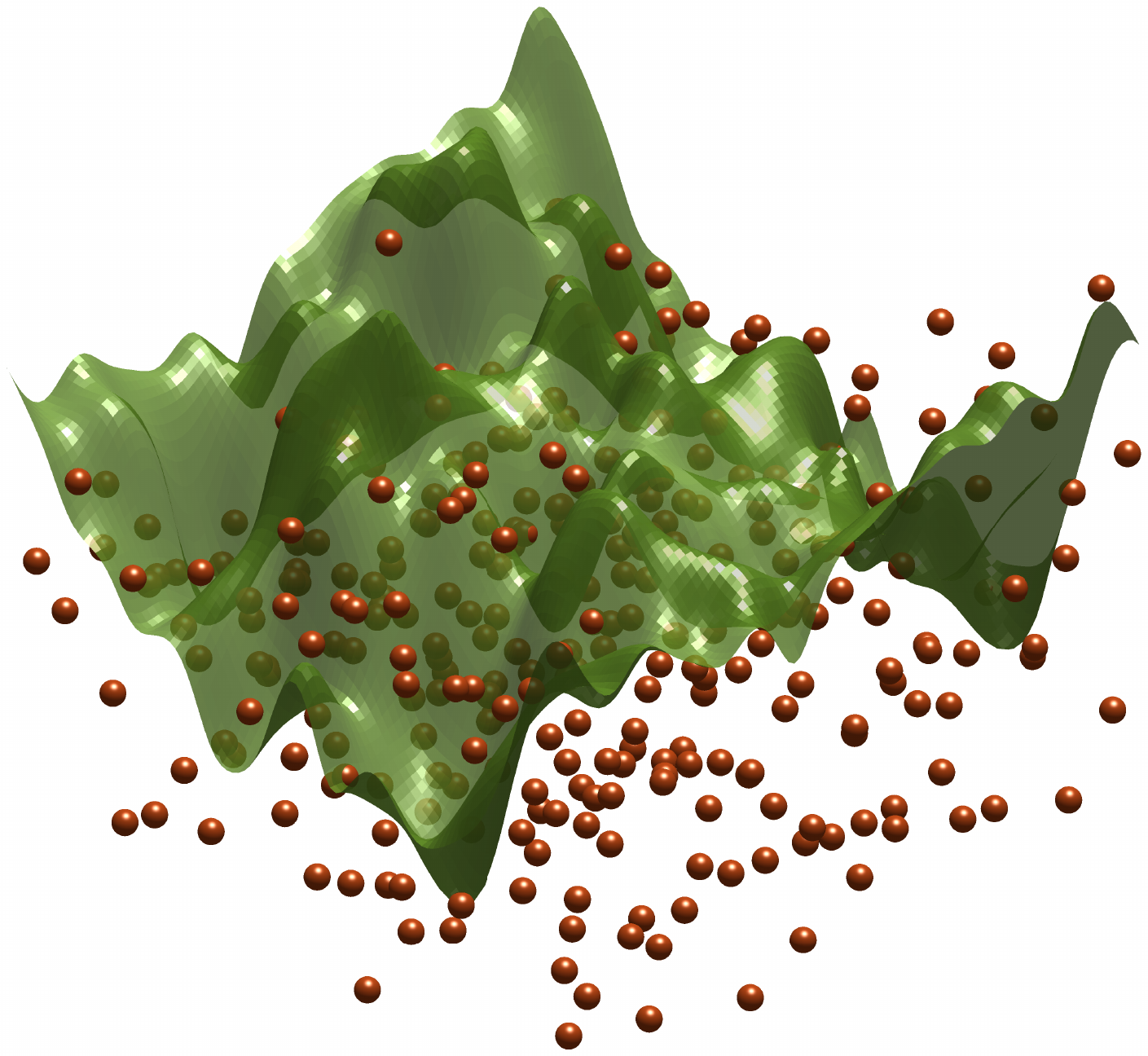}  & \textbf{\scriptsize (b)} & \includegraphics[scale=0.25, trim=20 100 0 0,clip, valign=t]{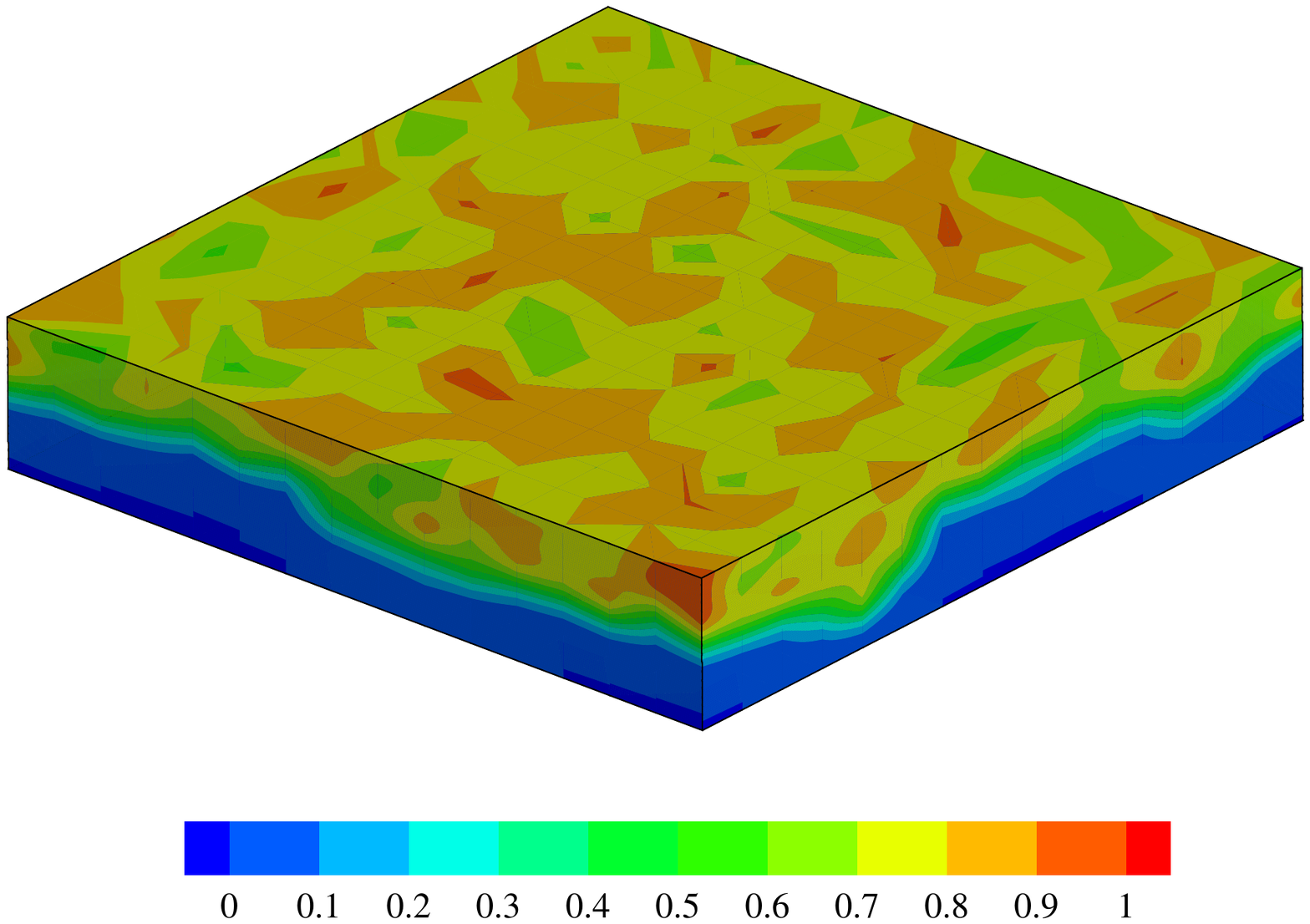} \\ 
\textbf{\scriptsize (c)} & \includegraphics[scale=0.2, valign=t]{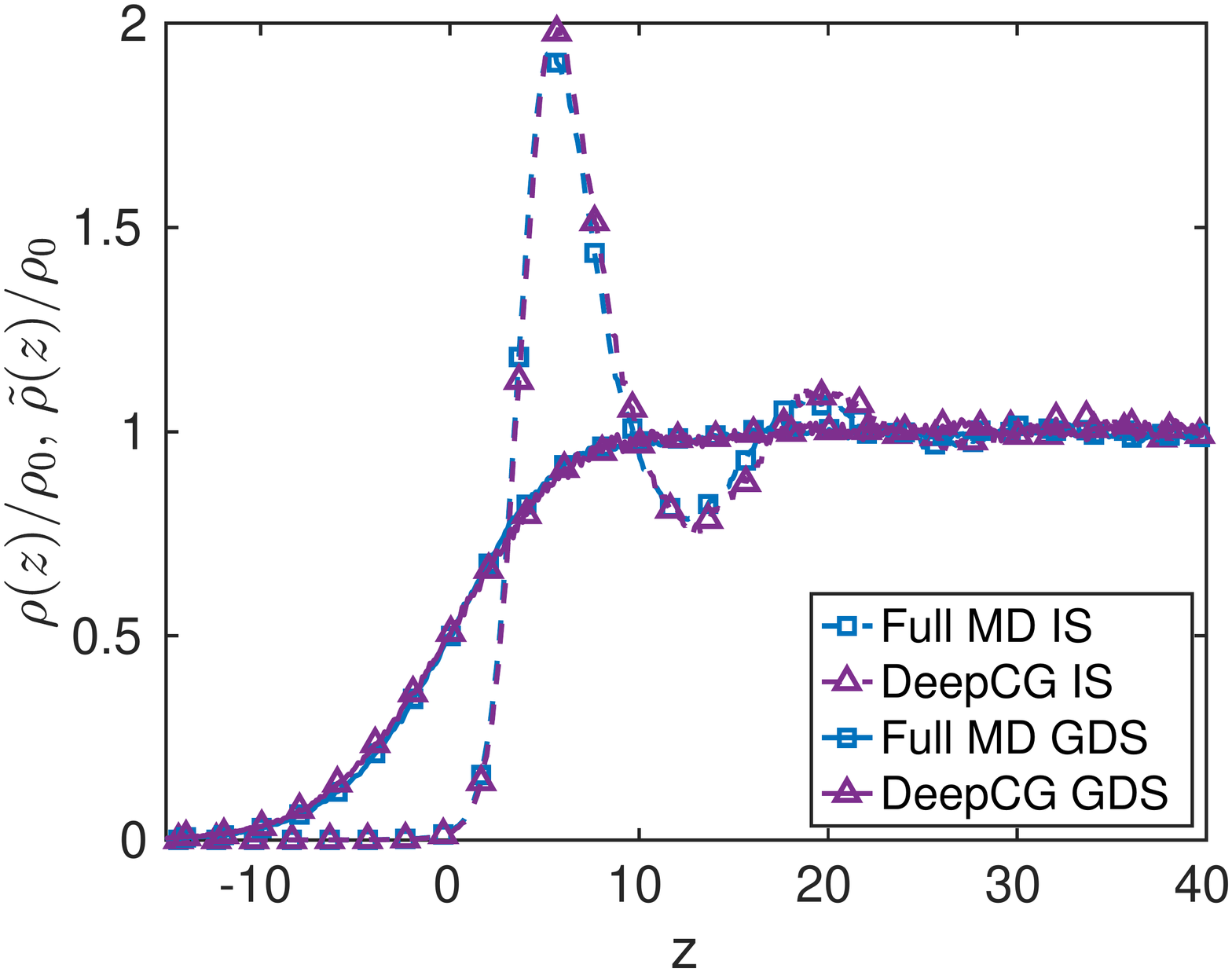} & \textbf{\scriptsize (d)} & \includegraphics[scale=0.2, valign=t]{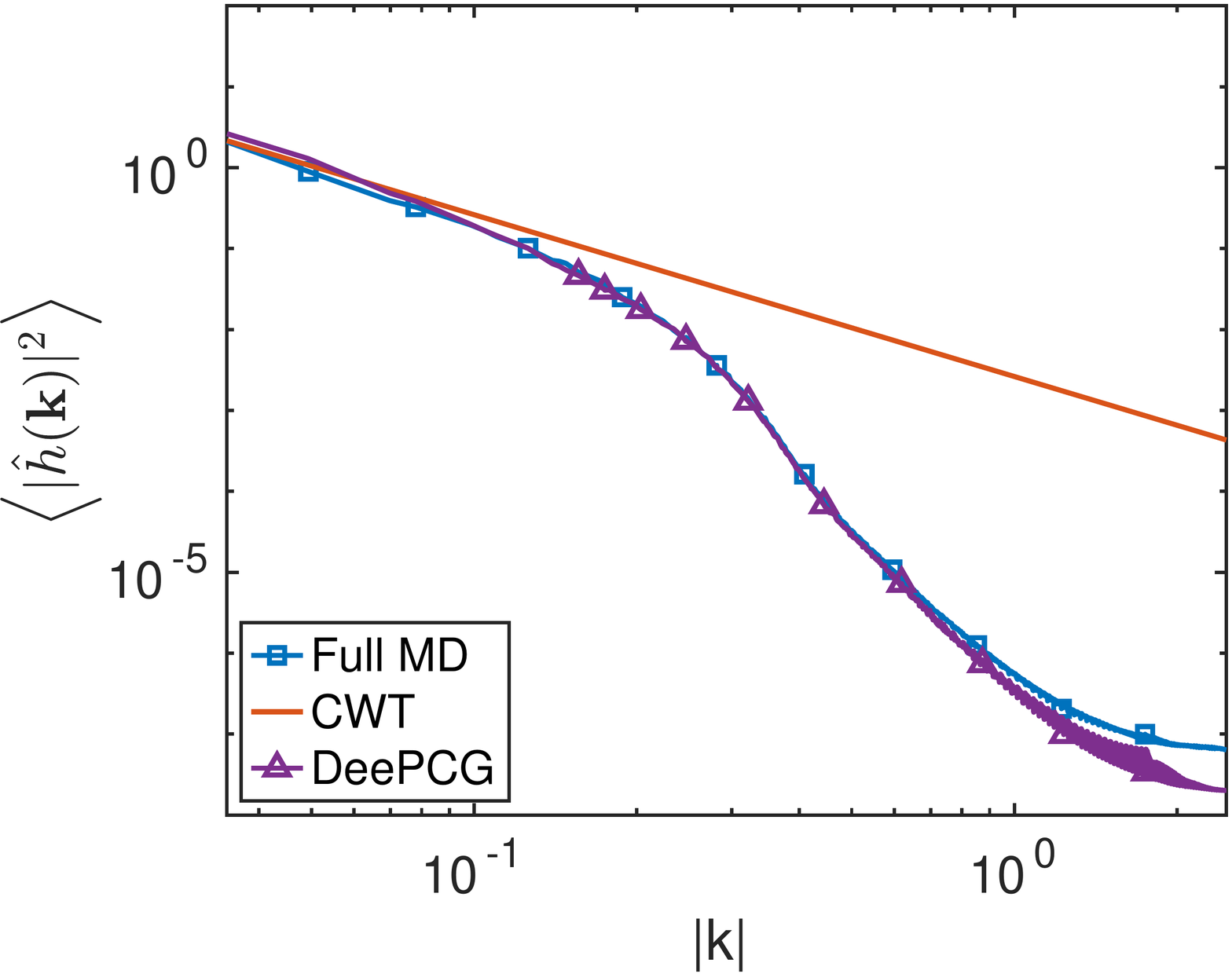}
\end{tabular}
\caption{The fluid density and the fluctuating interface of the single-component interfacial fluid system. 
(a) The interface defined by Eq. \eqref{eq:density_iosurf} with molecules (red) and interface (green). 
(b) A sketch of the instantaneous density field defined by Eq. \eqref{eq:density_field}. 
(c) The average density profile across the Gibbs dividing surface (GDS) and the instantaneous interface (IS) defined by
Eq. \eqref{eq:density_iosurf}. (d) The ensemble average of the capillary wave spectrum of the fluctuating interface. The solid line in red represents the CWT fitting using Eq. \eqref{eq:CWT} at the low wave number.}
\label{fig:single_component_interfacial_fluctuation}
\end{center}
\end{figure}

The smoothed density field $\rho_s(\mb R)$ enables us to define the instantaneous surface (IS) height $\tilde{h}(x,y)$ as the iso-surface of the fluid density \cite{Willard_Chandler_JPCB_2010}, i.e., 
\begin{equation}
\rho_s(x,y,\tilde{h}(x,y)) = \rho_0/2,
\label{eq:density_iosurf}
\end{equation}
where $\rho_0$ is the bulk fluid density, as shown in Fig. \ref{fig:single_component_interfacial_fluctuation}(a). Accordingly, we can compute the IS density distribution $\tilde{\rho}(z)$ along the $z$-direction, where the reference position is chosen to be $\tilde{h}(x,y)$ for each grid point $(x,y)$. As shown in Fig. \ref{fig:single_component_interfacial_fluctuation}(c), $\tilde{\rho}(z)$ exhibits apparent oscillations across the  instantaneous surface. The peaks near $z = 6$ and $z=16$ represent the first and the second layer of the fluid molecule near the interface. Alternatively, we can compute the density distribution $\rho(z)$ with respect to the plane at the average of the instantaneous height $\langle \tilde{h}(x,y)\rangle$, i.e., the Gibbs dividing surface. Different from $\tilde{\rho}(z)$,  $\rho(z)$ shows a smooth transition from $0$ to the bulk value across the interface. For both definitions, the predictions from the CG model agree well with the full MD simulations. We emphasize that the learning of the DeePCG potential \emph{does not} involve any human intervention such as the definitions of the density field and the interface height. The consistent predictions between the MD and CG models validate that constructed DeePCG potential $\tilde{U}(\mb Q; \theta)$ faithfully captures the intrinsic fluid structure near the interface.

To further examine the interfacial fluctuations, we evaluate the Fourier spectrum of the instantaneous height $\tilde{h}(x,y)$, i.e.,
\begin{equation}
\hat{h}(\mb k) = \frac{1}{L^2}\int_0^L\int_0^L \tilde{h}(x,y)e^{-ik_x x -i k_y y} \diff x \diff y,
\label{eq:density_height}
\end{equation}
where $\mb k = (k_x, k_y)$ is the 2D wave number. Fig. \ref{fig:single_component_interfacial_fluctuation} shows the ensemble average of the 
spectrum $\left\langle  \vert \hat{h}(\mb k)\vert^2\right\rangle$. On the low wave number limit, the interfacial energy 
is governed by the surface tensor with equi-partition distribution among the individual Fourier modes following the capillary wave theory (CWT) \cite{Buff_Lovett_PRL_1965, Evans_ADP_1979},  i.e., 
\begin{equation}
 \left\langle \left \vert \hat{h}(\mb k)\right\vert^2\right\rangle =\frac{k_BT L^2}{ \gamma {\left\vert \mb k\right\vert}^2}, 
 \label{eq:CWT}
\end{equation}
where $\gamma$ is the surface tension. At low wave number, 
$\left\langle  \vert \hat{h}(\mb k)\vert^2\right\rangle$ obtained from numerical simulations shows good agreement with the CWT theory.
As the wave number increases, the spectrum gradually deviates from the CWT prediction, indicating that there exists strong correlations between the height fluctuations of neighboring sites on the molecular scales. Nevertheless, the predictions 
from the CG model agree well with the MD results over the entire wave number regime. In particular, the good agreement in the high wave number regime shows that the CG model can accurately capture the local roughness of the interface, which is extremely sensitive to the molecule spatial correlations and the many-body interactions.

\begin{figure}[htbp]
\begin{center}
\begin{tabular}{ c c c c }
\textbf{\scriptsize (a)} & \includegraphics[scale=0.2, valign=t]{./void_probability_no_letter.pdf}  & \textbf{\scriptsize (b)} & \includegraphics[scale=0.2, valign=t]{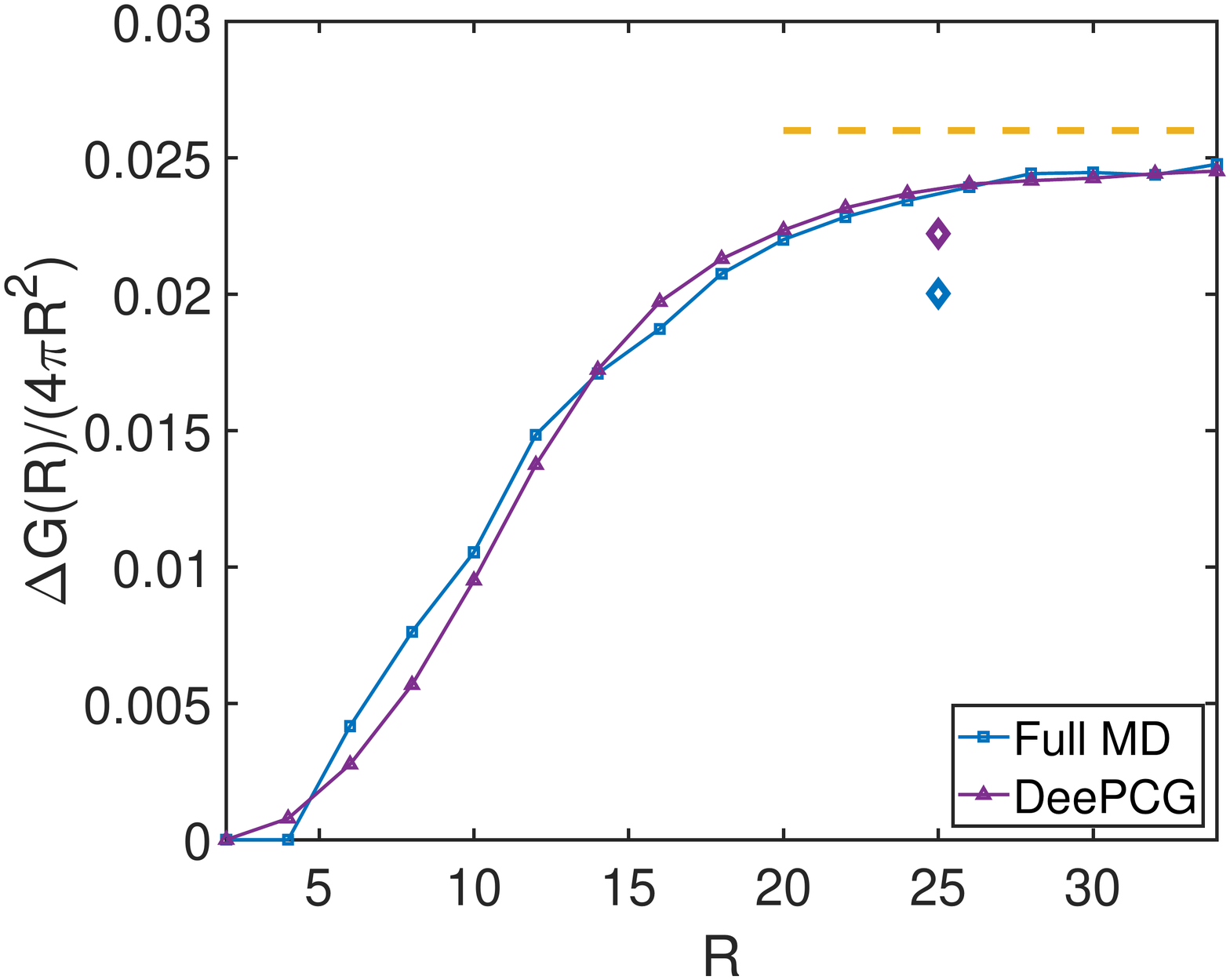}
\end{tabular}
\caption{(a) The probability density function of the molecule number within a spherical volume of radius $R = 25.0$. The red line represents the quadratic fitting; the deviation near $n=0$ arises from the formation of a clear fluid-void interface, where free energy approximately scales with the area of the interface. (b) Normalized solvation free energy $\Delta G(R)/4\pi R^2$ obtained from the thermal integration sampling by Eq. \eqref{eq:thermal_integration}. The transition from the volume- to area-scaling occurs between $R=15$ and $25$. The two symbols represent the predictions from the probability of the void space $-k_BT\ln P(0)$ for $R = 25$ in (a). The dashed horizontal line represents the macro-scale limit with the surface tensor $\gamma$ obtained from the fluctuating interface using CWT (Eq. \eqref{eq:CWT}) presented in Fig. \ref{fig:single_component_interfacial_fluctuation}. }
\label{fig:solvation_energy}
\end{center}
\end{figure}

Next, we examine the meso-scale, size-dependent apolar solvation energy. Similar to the bulk system considered in Sec. \ref{sec:bulk_fluids}, we examine the probability density function of the number of molecule $P(\hat{n})$ within a spherical volume of radius $R = 25.0$. As shown in Fig. \ref{fig:solvation_energy}(a), the predictions from the full MD and the CG model agree well over the full regime of $\hat{n}$. In particular, at the quasi-equilibrium regime, the interfacial energy is mainly determined by the fluid compressibility; $P(\hat{n})$ and $\hat{n}$ follow the quadratic relationship, i.e., 
$P(\hat{n}) \propto (\hat{n} - \langle n \rangle)^2/ \delta n^2$. Since both $\hat{n}$ and $\delta n^2$ scale with the volume, the free energy $-k_BT \ln P(\hat{n})$ scale with the volume near $\langle n \rangle$. 
In contrast, $P(\hat{n})$ deviates from the quadratic relationship as $\hat{n}$ decreases and yields a larger value of $P(0)$. The fat tail arises from the formation of a clear void-fluid interface. In particular, on the scale beyond the correlation length of fluid molecules, the local molecular reorganization is insufficient to accommodate the phase separation. Accordingly, the interfacial energy scales with the surface area of the void space. 

The multi-faceted nature of the interface energy can be further examined by computing the apolar solvation free energy $\Delta G = -k_BT \ln P(0)$ for the different sizes of the void space. By the theory of Pratt and his co-worker \cite{Hummer_Pratt_PNAS_1996}, for the small void space, $\Delta G$ is governed by the molecule number fluctuations with the Gaussian distribution, i.e., 
\begin{equation}
\Delta G\approx\frac{1}{2} k_BT \hat{n}^2 /  \delta n^2+ \frac{1}{2} k_BT \ln2\pi\delta n^2,  
\end{equation}
where $\hat{n}^2 / \delta n^2$ scales with the space volume $4/3 \pi R^3$. On the large scale, $\Delta G$ is determined by the macro-scale surface tensor $\gamma$, i.e., $\Delta G \approx 4\pi R^2 \gamma$. 

To quantify the cross-over regime, we conduct the thermal integration sampling of $\Delta G(R)$ with $R$ between $0$ and $34$. The integration force $\frac{\diff \Delta \tilde{G}}{\diff R}$ is estimated by imposing the biased potential, i.e.,
\begin{equation}
\frac{\diff \Delta G(R)}{\diff R}=\left\langle \sum_{i=1}^{M}  \nabla_{\mb Q_i} U_{bias} \cdot \frac{\mb Q_i - \mb Q_c}{\Vert \mb Q_i - \mb Q_c\Vert } \right\rangle,    
\label{eq:thermal_integration}
\end{equation}
where $U_{\rm bias}(\hat{n}; 0)$ is defined by Eq. \eqref{eq:bias_potential} with $k_n = 29.20$ and $h = 0.4$. Fig. \ref{fig:solvation_energy}(b) shows the obtained solvation energy $\Delta G(R)$ normalized by the surface area. The predictions of the CG and the full MD models show good agreement. In particular, at small value of $R$, $\Delta G(R)/4 \pi R^2$ grows with $R$ and implies the volume-scaling regime. The transition from the volume- to the area-scaling occurs between $R = 15$ and $R = 25$. For $R > 30$, $\Delta G(R)/4\pi R^2$ approaches the value of the macro-scale surface tensor $\gamma$ estimated from the interfacial fluctuations by the CWT theory \eqref{eq:CWT} shown in Fig. \ref{fig:single_component_interfacial_fluctuation}. 

\begin{figure}[htbp]
\begin{center}
\begin{tabular}{ c c c c }
\textbf{\scriptsize (a)} & \includegraphics[scale=0.2, valign=t]{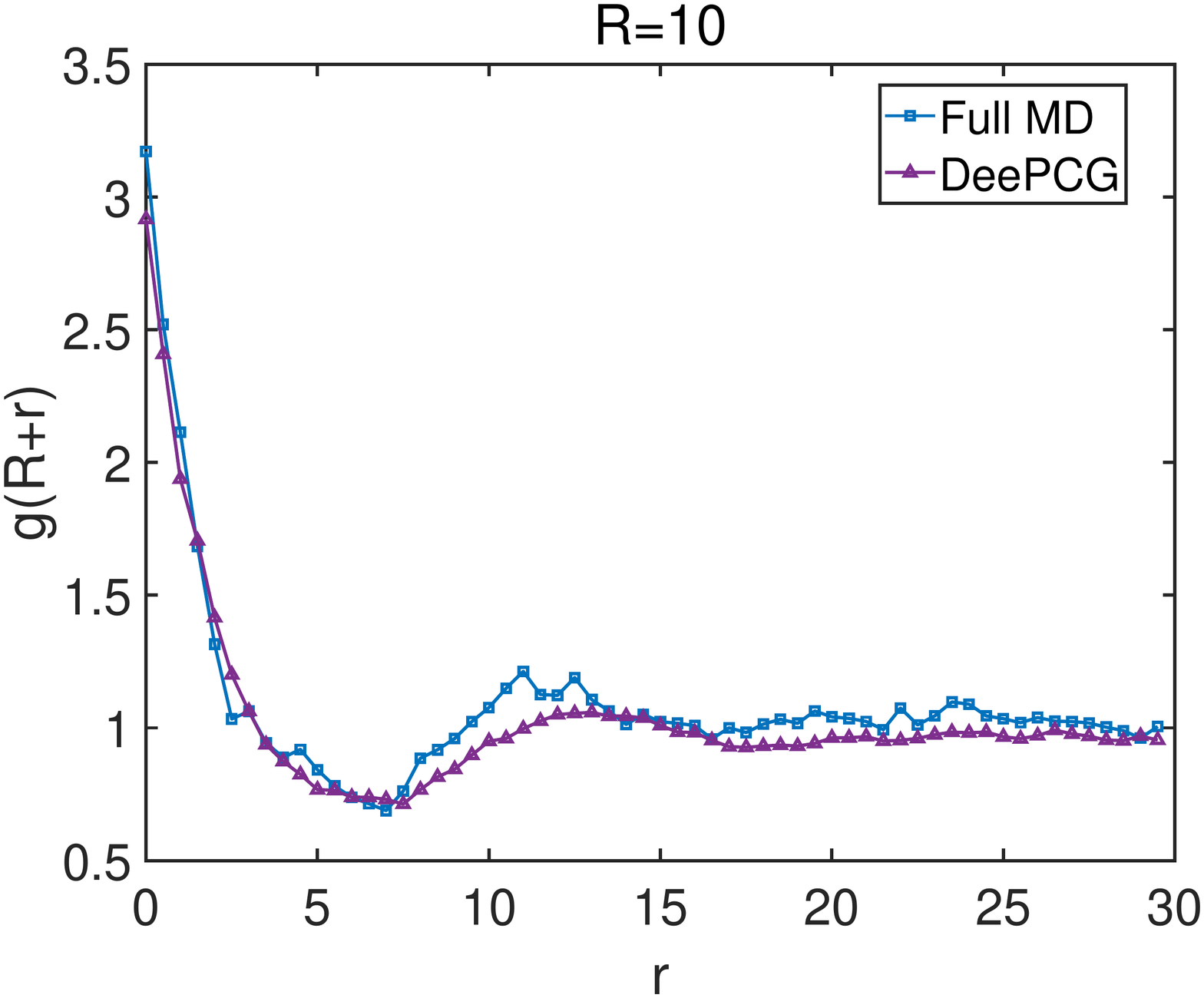}  & \textbf{\scriptsize (b)} & \includegraphics[scale=0.2, valign=t]{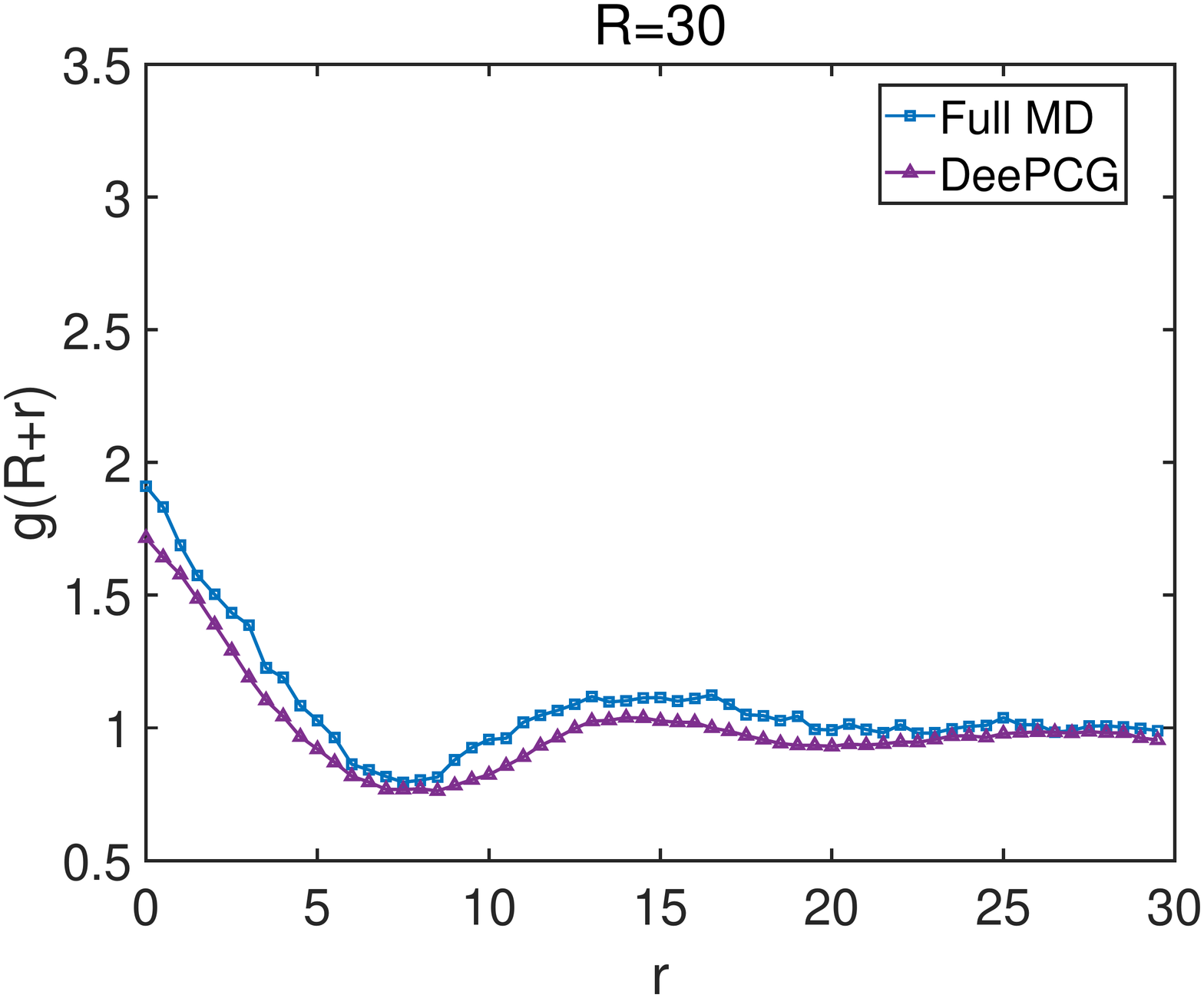}
\end{tabular}
\caption{The average equilibrium fluid density with a distance $r+R$, where $R$ is the radius of the spherical void space with $R = 10$ (left) and $R = 30$ (right).}
\label{fig:rcavity_pdf}
\end{center}
\end{figure}

The scale-dependent interfacial energy is also manifested in the solvent density distribution near the vicinity of the void space. Fig. \ref{fig:rcavity_pdf} shows the normalized radial distribution function $g(r+R)$ adjacent to the interface. For $R = 10$, solvation is governed by the local compressibility and molecule re-organization, leading to the high fluid density adjacent to the interface. For $R = 30$, solvation leads to the clear fluid-void interface and fluid density is closer to the bulk value. The CG model accurately captures the transition and agrees well with the full MD results for both cases.

\subsection{Two-component fluids}
\label{sec:two_component_fluids}
\begin{figure}[htbp]
\begin{center}
\begin{tabular}{ c c c c }
\textbf{\scriptsize (a)} & \includegraphics[scale=0.2, valign=t]{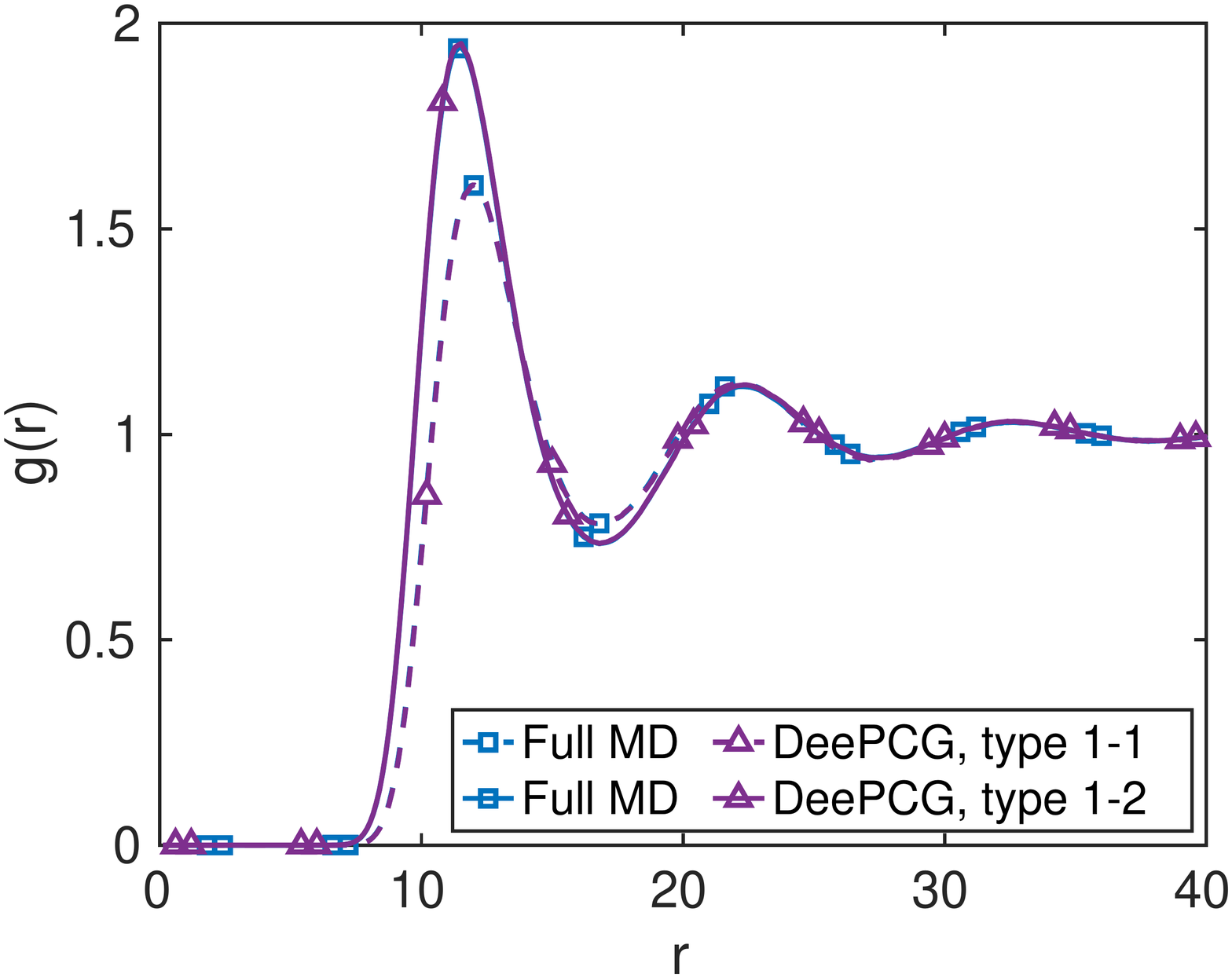}  & \textbf{\scriptsize (b)} & \includegraphics[scale=0.2, valign=t]{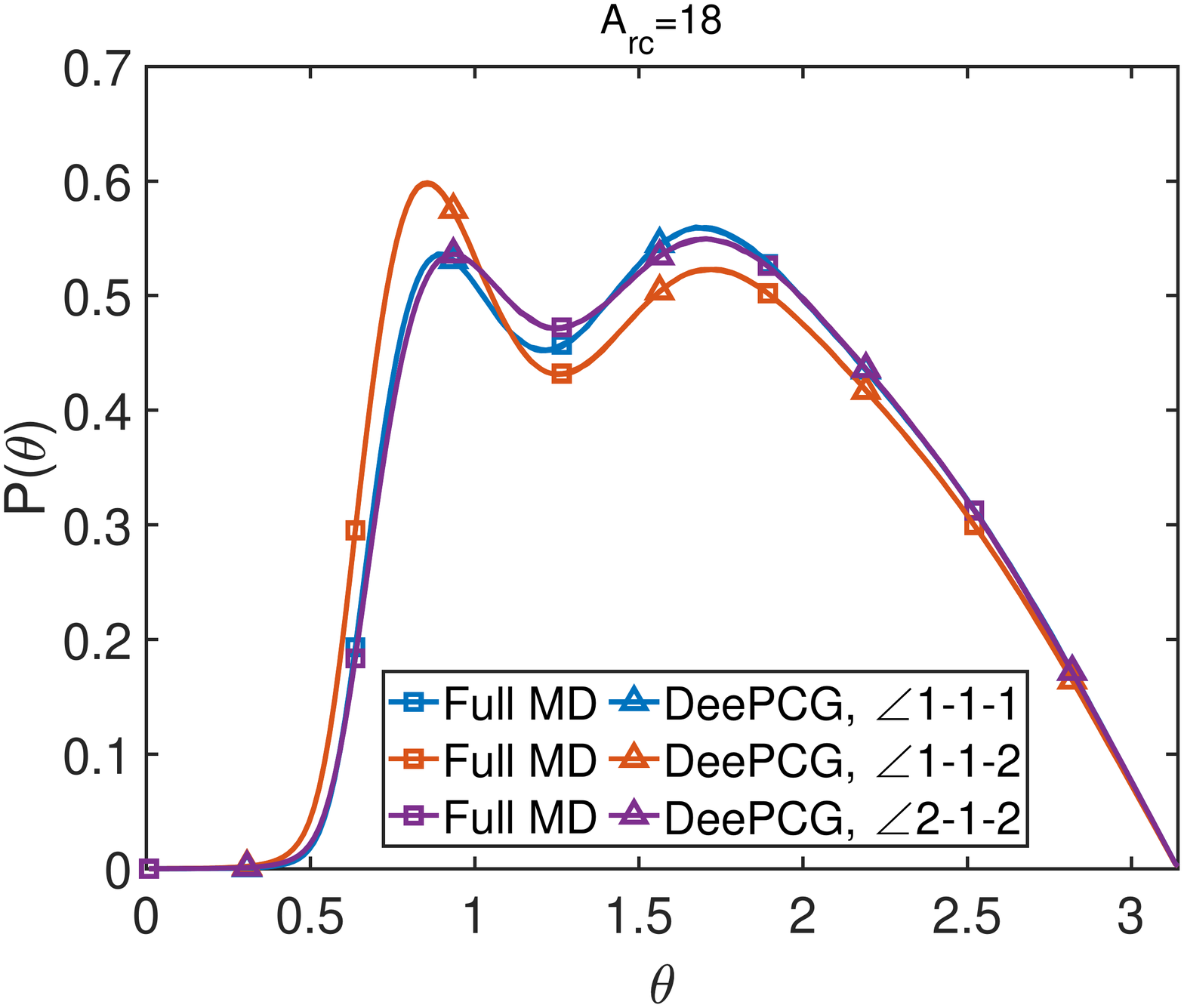}
\end{tabular}
\caption{(a) Radial distribution function $g(r)$ of the two-component, miscible polymer fluid system among the molecule COM of type 1-1, type 1-2. (b) Angular distribution function $P(\theta)$ of the same system among the molecule COM of type 1.
}
\label{fig:rdf_two_component}
\end{center}
\end{figure}

We first consider a two-component fluid system that takes the parameter set (\Rmnum{1}) specified in Sec. \ref{sec:full_model}. 
The potential between type$\mhyphen$1
and type$\mhyphen$2 molecules is lower than the value of the single component. Therefore, the full MD system can maintain a full mixture state. The reduced model
is represented by the CG particles of two different types. The equilibrium state reaches a full mixture state as well.
Fig. \ref{fig:rdf_two_component} shows the radial distribution functions of the COM of the molecules.
Due to the ``hydrophilic'' interactions between type-1 and $\mhyphen$2 molecules,
the pair distribution between type 1-2 shows more a pronounced peak at $R = 11.5$ as compared with the distribution between type 1-1 at $R = 12.5$.  Similar to Sec. \ref{sec:bulk_fluids}, we compute the angular distribution functions among the molecules of both types. For all of the correlation functions, the CG and full MD models show good agreement.

\begin{figure}[htbp]
\begin{center}
\begin{tabular}{ c c c c}
\textbf{\scriptsize (a)} & \includegraphics[scale=0.2, trim=0 20 20 0,clip, valign=t]{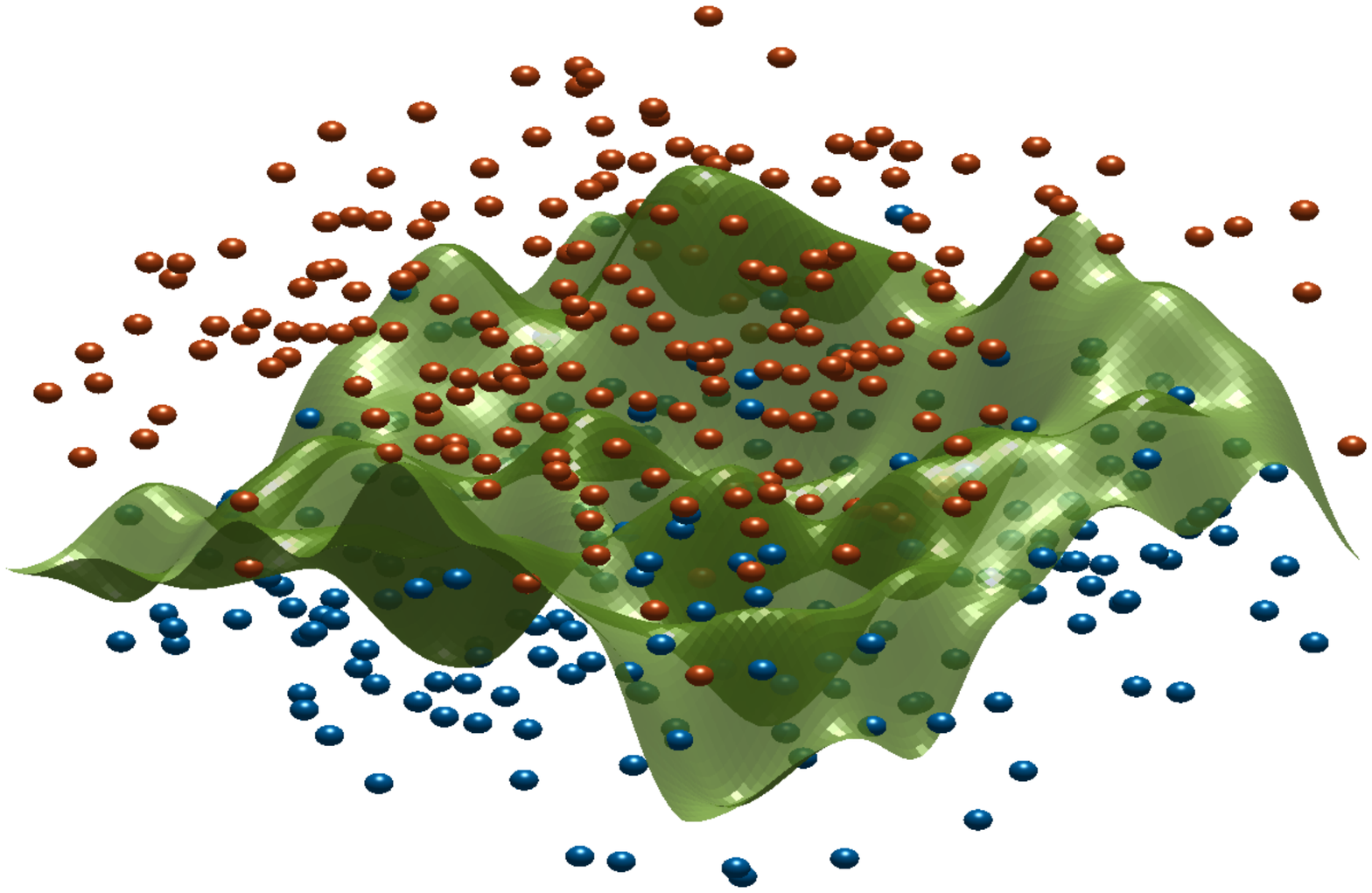} & \textbf{\scriptsize (b)} & \includegraphics[scale=0.2, valign=t]{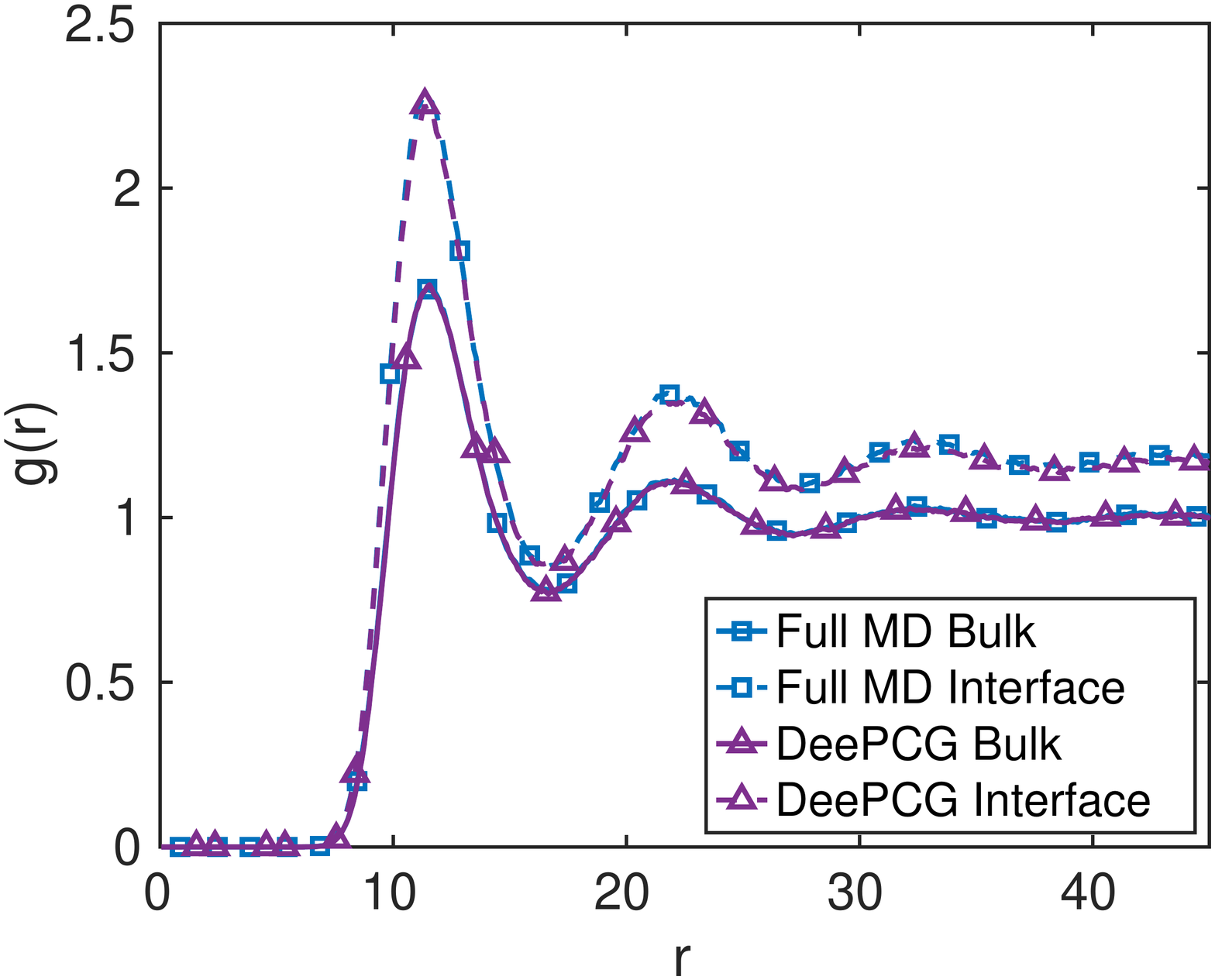}  \\ 
\textbf{\scriptsize (c)} & \includegraphics[scale=0.2, valign=t]{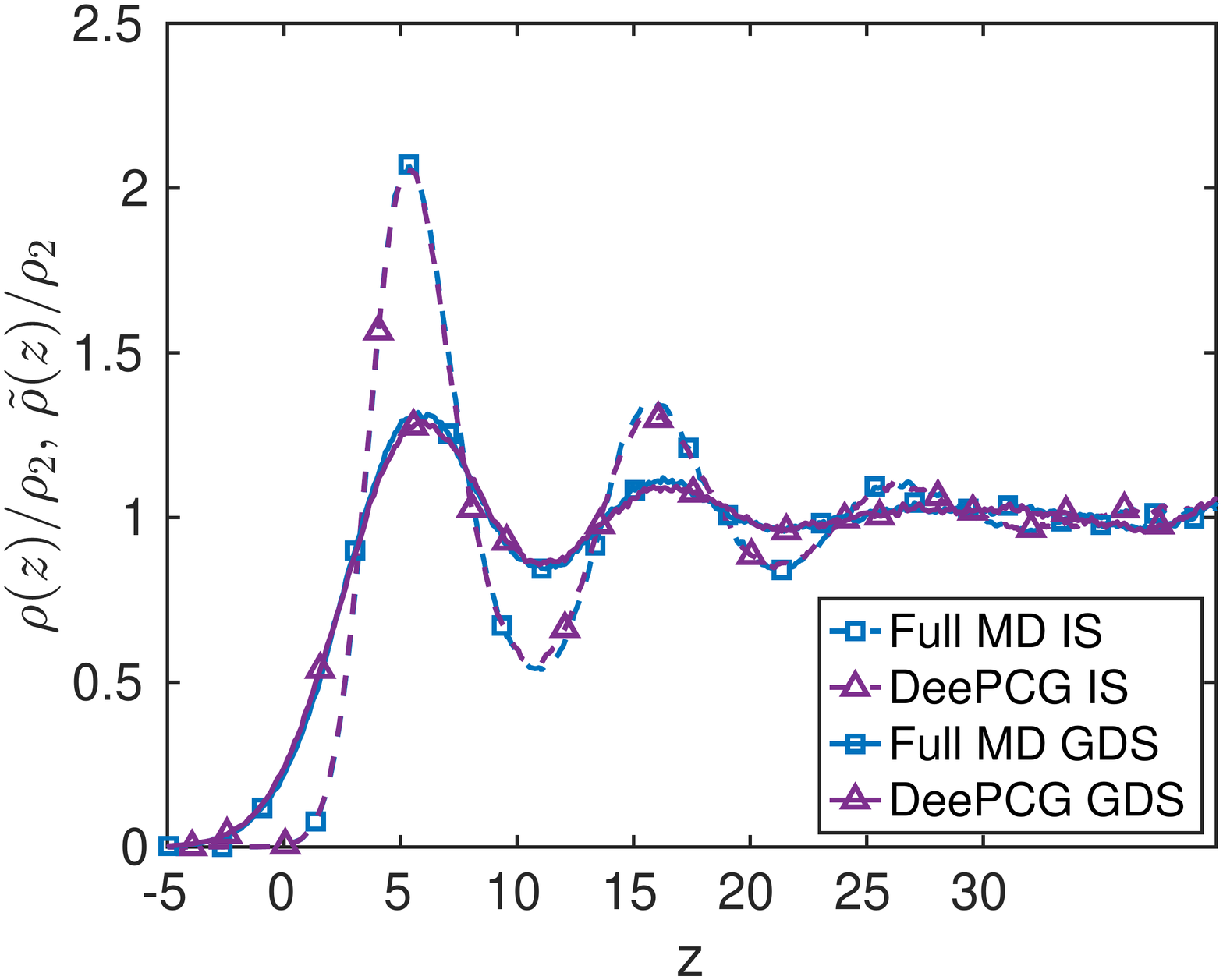} & \textbf{\scriptsize (d)} & \includegraphics[scale=0.2,  valign=t]{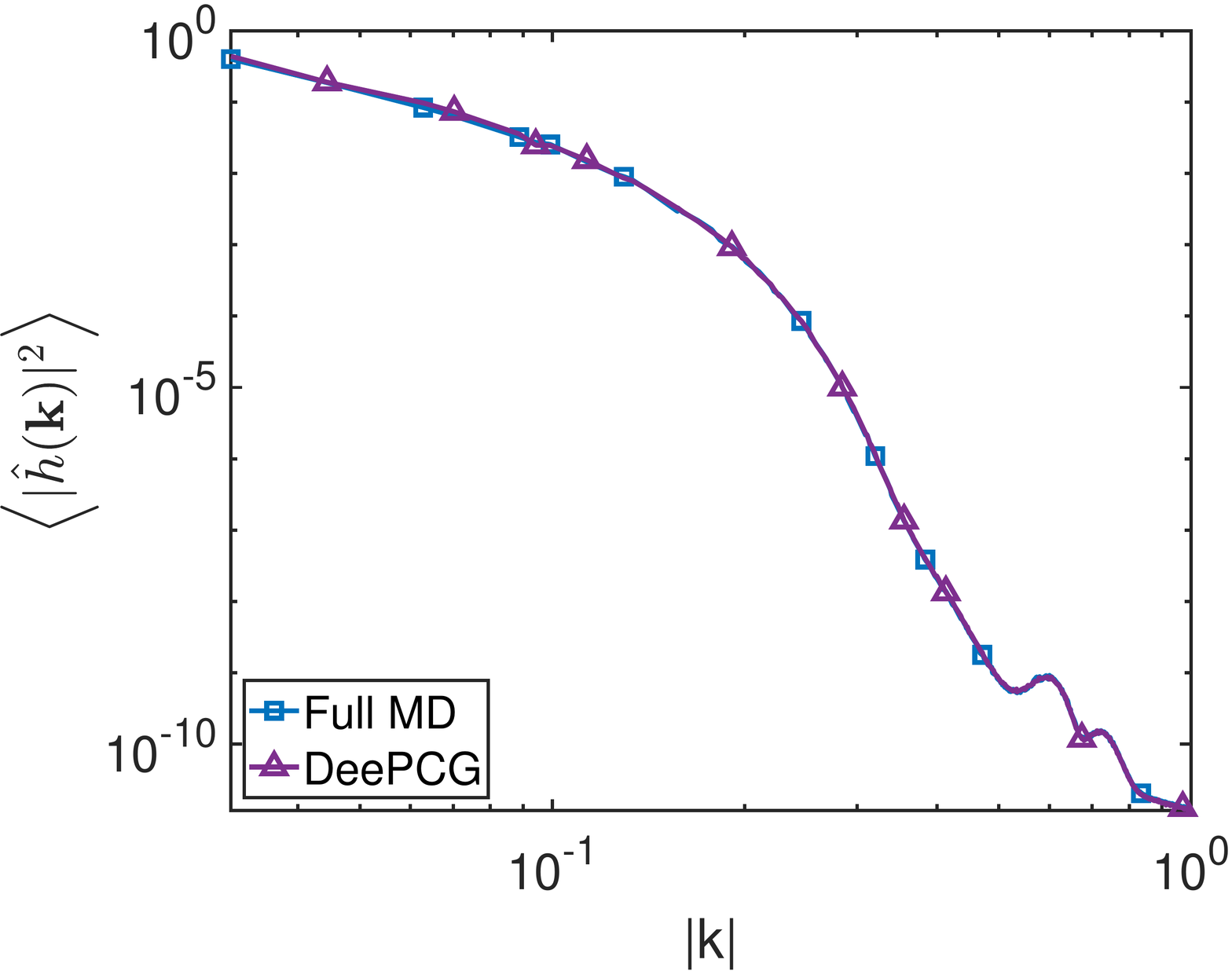} 
\end{tabular}
\caption{The fluid density and the fluctuating interface of the two-component, immiscible fluid system. 
(a) The interface defined by Eq. \eqref{eq:density_height} with type-1 (blue) and type-2 molecules (red). 
(b) Radial distribution function $g(r)$ of type-2 molecules on the $x\mhyphen y$ plane near the bulk ($z = 30$) and the interface ($z = 60$). 
(c) The average density profile across the Gibbs dividing surface  and the instantaneous surface defined by
Eq. \eqref{eq:density_iosurf}. (d) The capillary wave spectrum of the fluctuating interface.}
\label{fig:interfacial_fluctuation}
\end{center}
\end{figure}

Next, we consider the parameter set (\Rmnum{2}) specified in Sec. \ref{sec:full_model}. Due to the ``hydrophobic'' interaction between the two molecule types, the
system develops into an immiscible state with a clear interface between the two components, as shown in Fig. \ref{fig:interfacial_fluctuation}(a). To examine the heterogeneous fluid particle distribution, we analyze the radial distribution functions of the fluid particle on the $x\mhyphen y$ plane at different regimes.  Fig. \ref{fig:interfacial_fluctuation}(b) shows the planar RDFs sampled at $z = 60$ (interface) and $z = 30$ (bulk).  In particular, the planar RDF near the interfacial regime shows more pronounced peaks and structural oscillations compared with the RDF in the bulk regime. For both cases, the predictions from the CG model show good agreement with the full MD simulations.

To further quantify the fluid density across the interface, we define the density field $\rho_s(\mb R)$ by Eq. \eqref{eq:density_field} 
on the lattice grids across the average interface of the two components (i.e., GDS) and the instantaneous height $\tilde{h}(x,y)$ as the iso-surface of the fluid density of a single component (i.e., IS). For this system, we set $h = 40$, $dl = 2.0$ and $dz = 1.0$. 
Fig. \ref{fig:interfacial_fluctuation}(c) shows the density profiles
$\tilde{\rho}(z)$ and $\rho(z)$ across the interface based on the 
definition of IS and GDS, respectively. Similar to the single-component fluid system, $\tilde{\rho}(z)$ shows pronounced oscillations that represent the intrinsic multi-layer fluid structure across the interface. In contrast, $\rho(z)$ shows a smooth transition across the interface due to the ensemble-averaged definition of the interface plane. The consistent predictions between the MD and CG models validate the accuracy of the constructed DeePCG potential.

Finally, we examine the thermal fluctuations across the interface. Fig. \ref{fig:interfacial_fluctuation}(d) shows the ensemble average of the Fourier spectrum density $\left\langle \vert \hat{h}(\mb k)\vert^2\right\rangle$ of the instantaneous height $\tilde{h}(x,y)$ defined by Eq. \eqref{eq:density_height}. Similar to the single-component interfacial fluid system, $\left\langle \vert \hat{h}(\mb k)\vert^2\right\rangle$ agrees well with the CWT theory at the low wave number and deviates from the $1/\vert \mb k \vert^2$ scaling at high wave number due to the local spatial correlations between the molecules. The predictions from CG and full MD models show good agreement over the full regime.

\section{Summary}
\label{eq:summary}
In this study, we constructed coarse-grained models of meso-scale interfacial polymeric fluids based on the DeePCG scheme \cite{Zhang_DeePCG_JCP_2018}. In particular, the constructed CG potential can accurately encode the many-body interactions arising from the unresolved atomistic interactions, as well as the heterogeneous molecule distributions near the interface. This unique feature ensures that the constructed CG models can retain the consistent invariant distribution with the full
MD model and faithfully capture the multi-facted, scale-dependent interfacial energy without additional human intervention. The training process only requires the MD samples of the instantaneous force field without further \emph{ad hoc} assumptions and approximations of the CG potential functions.

While we focus on the polymeric fluids in this study, the present CG models can be generalized for complex fluids and soft matter systems where the many-body and heterogeneous effects are often pronounced. In particular, the constructed CG potential functions accurately reproduce the pairwise and high-order correlation functions while the empirical approximations show limitations. Moreover, the accurate predictions of the local compressibilty and the full-range spectrum of the interfacial fluctuations demonstrate the validity of the CG models to probe the collective behaviors across the molecular and continuum scales.  
More importantly, the CG models successfully predict the probability of the void formation as a rare event and the transition of the volume$\mhyphen$ to area$\mhyphen$scaling of solvation energy.  The accurate predictions on such properties show the promise of the present models to study the challenging problems relevant to nanoscale assembly processes \cite{Miller_Eric_PNAS_2007}, where the full MD simulations often show limitation to achieve the resolved spatio-temporal scale.

Finally, we note that the present study focuses on the quasi-equilibrium properties of the reduced model. The predictive modeling of the dynamic properties further relies on the accurate construction of the memory and fluctuation terms that represent the unresolved energy-dissipation processes \cite{hijon2010mori,Lei_Li_PNAS_2016, Lei_Li_JCP_2021}. We will pursue this problem in future studies.


\begin{acknowledgments}
The work is supported by the Extreme Science and Engineering Discovery Environment (XSEDE) Bridges at the Pittsburgh Supercomputing Center through allocation MTH210005. PG and HL are partially supported by the National Science Foundation under Grant DMS-2110981 and CAREER award DMS-2143739. 
\end{acknowledgments}

%

\end{document}